\documentclass[aps,prc,showpacs,showkeys,superscriptaddress,reprint,groupedaddress,floatfix]{revtex4-2}

\usepackage{graphicx}
\usepackage{amsmath}
\usepackage{amssymb}
\usepackage{bm}
\usepackage{placeins}

\begin{document}
\title{Classification of fission modes in $^{236}$U using a six-dimensional Langevin approach}
\author{K. Okada}
\email{okada.kazuki@jaea.go.jp}
\affiliation{Advanced Science Research Center, Japan Atomic Energy Agency, Tokai, Ibaraki 319-1195, Japan}
\author{K. Nishio}
\affiliation{Advanced Science Research Center, Japan Atomic Energy Agency, Tokai, Ibaraki 319-1195, Japan}
\author{T. Wada}
\affiliation{Department of Pure and Applied Physics, Kansai University, Suita, Osaka 564-8680, Japan}
\author{N. Carjan}
\affiliation{
  Joint Institute for Nuclear Research, 141980 Dubna, Moscow Region, Russia}
\affiliation{
  University of Bordeaux, CNRS, LP2i Bordeaux, UMR 5797,
  F-33170 Gradignan, France}
\date{\today}

\begin{abstract}
	Thermal neutron-induced fission of $^{235}$U is studied using a six-dimensional
	Langevin approach based on the Cassini shape parametrization.
	Scission events are classified into Asymmetric 1 (AS1), Asymmetric 2 (AS2), and
	Superlong (SL) fission modes by applying the $k$-means algorithm to the fragment
	mass and the quadrupole deformations of both fragments.
	For each mode, proton and neutron single-particle levels are calculated for
	representative fragments to examine their shell structures. The AS1 heavy
	fragment exhibits proton gaps at $Z=50$ and 52 and neutron gaps at $N=82$ and
	84, whereas well-developed gaps appear at $Z=56$ and $N=88$ in the AS2 heavy
	fragment.
	The mass splits of AS1 and AS2 are close to those of the conventional
	Standard~I and Standard~II modes, respectively. However, the average total
	kinetic energy is lower for AS1 than for AS2, opposite to the conventional
	ordering of Standard~I and Standard~II.
	This reversal reflects the more elongated shape of the AS1 light fragment.
	The SL mode is conventionally interpreted in terms of macroscopic liquid-drop
	effects, whereas the pronounced proton shell gap at $Z=46$ suggests that proton
	shell effects also contribute to the elongated symmetric configuration.
	The classification based on fragment mass and the quadrupole deformations of
	both fragments provides a basis for distinguishing fission modes and examining
	the corresponding fragment shell structures at scission.
\end{abstract}

\maketitle
\section{Introduction}\label{sec:introduction}

Nuclear fission is a fundamental process in heavy nuclei. One of its characteristic
features is the asymmetry of the fragment mass distribution. In low-energy fission of
light actinides, the yield is largest for asymmetric mass division, and the heavy
fragment mass distribution is concentrated near
$A_H=140$~\cite{vandenbosch1973,wagemans1991,andreyev2018}. In the classical liquid-drop
model, the fission barrier arises from the competition between repulsive Coulomb and
attractive surface energies~\cite{bohr1939}, and the model predicts mass-symmetric
fission. Shell effects are therefore needed to explain asymmetric mass distributions.

The shell structure of the nucleus creates a complicated potential energy surface and
the double-humped fission barrier~\cite{strutinsky1967,strutinsky1968}. The nucleus can
therefore evolve along several paths toward different scission configurations. In the
conventional description of low-energy fission in light actinides, fragment mass and
total kinetic energy (TKE) distributions are decomposed into two asymmetric modes,
Standard I and Standard II, and the symmetric Superlong (SL) mode~\cite{brosa1990}. In
this description, the SL mode is characterized by relatively low TKE and an elongated
scission configuration. For thermal neutron-induced fission of $^{235}$U, Standard I
exhibits a less asymmetric mass division and a higher TKE than Standard
II~\cite{knitter1987,hambsch1989}. For the Standard I and Standard II modes, the heavy
fragment proton numbers are approximately $Z_H=52$ and 56, respectively%
~\cite{martin2021,berriman2022,banerjee2023,dey2025}.

The shell structure of nascent fragments can be examined in microscopic calculations as
the fissioning system approaches scission. Heavy fragments in the $^{144}$Ba region were
found to exhibit octupole deformation in time-dependent Hartree-Fock calculations with
dynamical Bardeen-Cooper-Schrieffer pairing correlations (TDBCS)~\cite{scamps2018}. This
deformation is stabilized by shell gaps at $Z=52,56$ and $N=84,88$ and plays an
important role in determining the fission mass asymmetry. In constrained
Hartree-Fock-Bogoliubov (HFB) calculations for $^{236}$U, related shell gaps were also
found in the heavy fragment, and a proton shell gap at $Z=38$ was identified in the
light fragment for the prefragment configuration
$^{96}$Sr+$^{140}$Xe~\cite{bernard2023}. The octupole-deformed heavy fragments near
$Z=56$ correspond directly to Standard II, whereas the $Z=52$ shell gap is associated
with Standard I. For Standard II, both fragments are deformed, with the light fragment
more strongly deformed than the heavy fragment. However, it remains difficult to compare
these microscopic results directly with experimental fragment mass and TKE
distributions.

Among theoretical approaches, the multidimensional Langevin equation has been widely
used to calculate fragment mass and TKE
distributions~\cite{wada1993,karpov2001,aritomo2013,usang2016,ishizuka2017,sierk2017,usang2017,usang2019,liu2021}.
An appropriate shape parametrization is needed to represent fragment shapes effectively.
Recently, a five-dimensional Langevin calculation based on the two-center shell model
(TCSM) was used to describe the transition from mass-symmetric to mass-asymmetric
fission in Th isotopes~\cite{ivanyuk2024b}. This framework has also been used to
calculate fragment mass and TKE distributions for Hg fission~\cite{ivanyuk2025}. These
studies have compared the calculated distributions with experimental data, but detailed
analyses of fragment shapes and their shell structures at scission remain
limited.

In the present study, we employ a six-dimensional Langevin calculation based on the
Cassini shape parametrization~\cite{pashkevich1971}. Whereas the TCSM was constructed to
describe specific deformations occurring during fission, the Cassini parametrization
describes nuclear shapes through a mathematical expansion, allowing a wide range of
nuclear deformations to be incorporated naturally into a multidimensional calculation.
It has also been used in the scission-point
model~\cite{carjan2015,carjan2017,carjan2017b,carjan2019}. In particular, it was found
that the shape of the fissioning nucleus near scission can be flexibly described using
six deformation parameters~\cite{ivanyuk2024c}. As a dynamical application, a
five-dimensional Langevin calculation has been performed for $^{236}$U~\cite{okada2025}.
Here, we examine whether the calculated fission events can be classified into
fission modes on the basis of the fragment mass and the quadrupole deformations
of both fragments.

The paper is organized as follows.
Section~\ref{sec:theory} describes the theoretical framework, including the
Cassini shape parametrization, the multidimensional Langevin equation, and the
mode classification procedure.
Section~\ref{sec:results} presents the decomposition of scission events into
fission modes, the mass--TKE distributions for each mode, and the
single-particle levels of representative fragments.
Section~\ref{sec:summary} summarizes the present study.

\section{Theoretical framework}\label{sec:theory}

\subsection{Cassini shape parametrization}\label{sec:cassini}

Nuclear shapes during fission are described by the Cassini shape
parametrization \cite{pashkevich1971}. In this parametrization, the nuclear
surface is generated from Cassini ovals expressed in lemniscate coordinates.
Compact, necked, and near-scission configurations can therefore be represented
in a common coordinate system. The surface is first specified in the scaled
cylindrical coordinates $(\bar{\rho},\bar{z})$, which are expressed in terms
of the lemniscate coordinates $(R,x)$ as
\begin{align}
  \bar{\rho}
  &= \frac{1}{\sqrt{2}}
  \sqrt{p(x)-R^2(2x^2-1)-s},
  \nonumber\\
  \bar{z}
  &= \frac{\mathrm{sign}(x)}{\sqrt{2}}
  \sqrt{p(x)+R^2(2x^2-1)+s},
  \nonumber\\
  p(x)
  &\equiv \sqrt{R^4+2sR^2(2x^2-1)+s^2},
  \nonumber\\
  &0 \leq R < \infty,
  \qquad
  -1 \leq x \leq 1.
  \label{eq:Cassini1}
\end{align}
Here, $s$ controls the elongation from a compact shape toward two separated
fragments.
The physical cylindrical coordinates $(\rho,z)$ are obtained from the scaled coordinates as
\begin{align}
  \rho &= \bar{\rho}/c,
  \nonumber\\
  z &= (\bar{z}-\bar{z}_{\mathrm{cm}})/c,
  \label{eq:Cassini2}
\end{align}
where $c$ is fixed by volume conservation and $\bar{z}_{\mathrm{cm}}$ is fixed
by requiring the center of mass to be at the origin of the $(\rho,z)$
coordinate system.

Additional deformation degrees of freedom are introduced by
expanding $R$ in a series of Legendre polynomials,
\begin{align}
  R(x)=R_0\left[1+\sum_n \alpha_n P_n(x)\right],
  \label{eq:Cassini3}
\end{align}
where $R_0$ is the radius of the corresponding spherical nucleus and $P_n(x)$
is the $n$th Legendre polynomial. The deformed nuclear shapes are
characterized by the expansion coefficients $\alpha_n$.

\begin{figure}[t]
  \centering
  \includegraphics[width=\columnwidth, trim=1.73cm 7.55cm 3.05cm 5.80cm, clip]{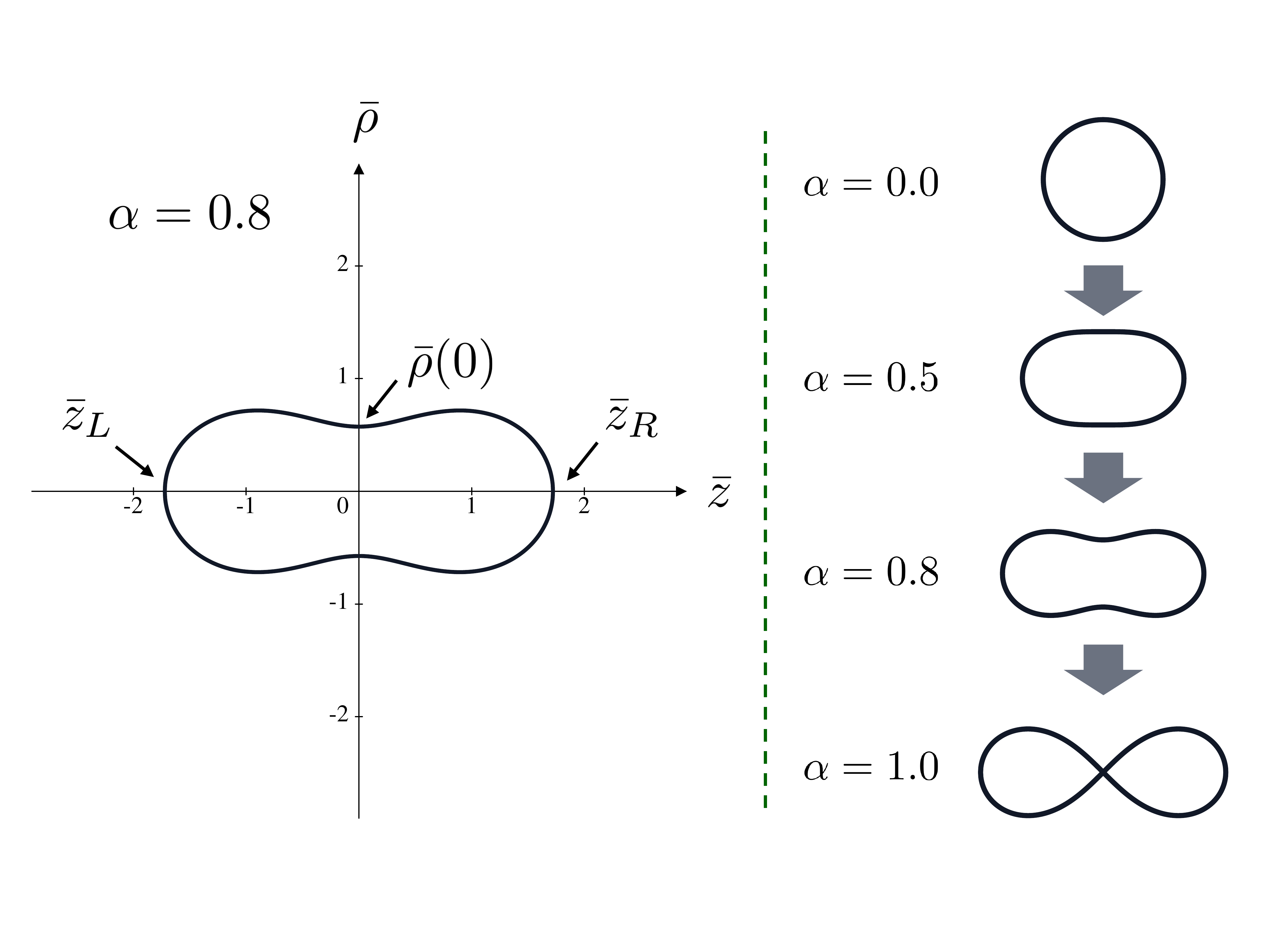}
  \caption{Elongation parameter $\alpha$ in the Cassini shape parametrization.
  The left panel shows the shape at $\alpha=0.8$ in the scaled cylindrical coordinates
  $(\bar{z},\bar{\rho})$ and indicates $\bar{z}_L$, $\bar{z}_R$, and $\bar{\rho}(0)$,
  which are used in Eq.~(\ref{eq:Cassini4}). The right panel shows shapes
  at $\alpha=0.0$, 0.5, 0.8, and 1.0, with $\alpha_n=0$ for all $n$.}
  \label{fig:cassini_alpha}
\end{figure}

As an alternative to $s$, the elongation parameter $\alpha$ is defined as
\begin{align}
  \alpha
  &\equiv
  \frac{\bar{z}_R^2+\bar{z}_L^2-2\bar{\rho}^2(0)}
  {\bar{z}_R^2+\bar{z}_L^2+2\bar{\rho}^2(0)}.
  \label{eq:Cassini4}
\end{align}
Here $\bar{z}_R$ and $\bar{z}_L$ are the right and left endpoints of the
nuclear surface on the symmetry axis in the scaled coordinates, and
$\bar{\rho}(0)$ is the radius at the midplane. These geometrical quantities
are illustrated in the left panel of Fig.~\ref{fig:cassini_alpha}.

The right panel of Fig.~\ref{fig:cassini_alpha} shows the dependence of the
shape on $\alpha$. With $\alpha_n=0$ for all $n$, the shape is spherical at
$\alpha=0$ and becomes elongated with a neck at the midplane as $\alpha$
increases. At
$\alpha=1$, the neck radius becomes zero. Although $s$ also controls the elongation, the neck radius at a given $s$
depends strongly on the deformation coefficients $\alpha_n$. Thus, $\alpha$
is used as the main elongation coordinate in the Cassini shape parametrization.

\begin{figure}[t]
  \centering
  \includegraphics[width=\columnwidth, trim=2.45cm 7.15cm 0.0cm 1.95cm, clip]{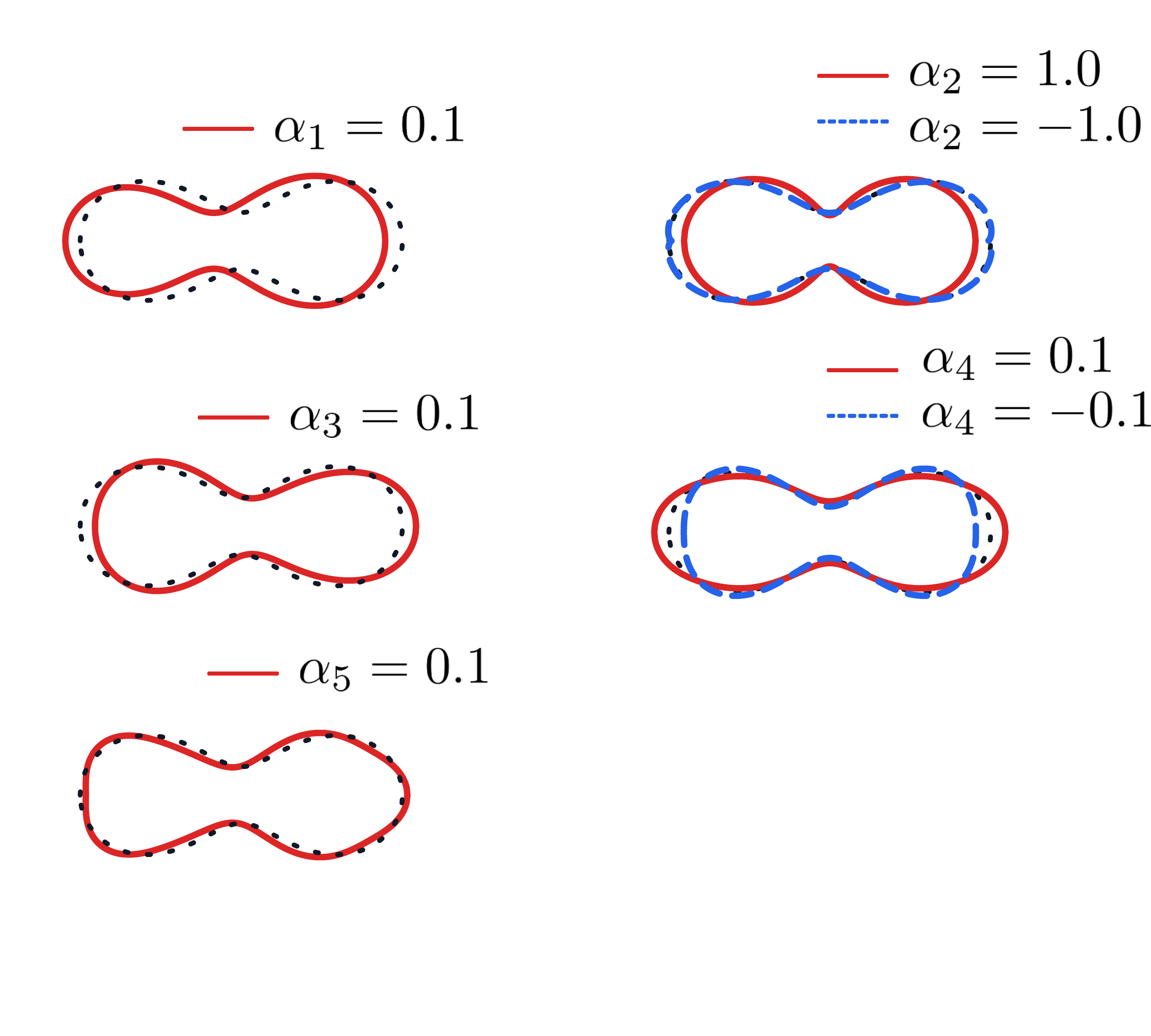}
  \caption{Geometrical effects of the deformation coefficients $\alpha_n$ at fixed elongation
  $\alpha=0.98$. In each panel, the black dotted and red solid curves show the
  reference shape with all $\alpha_n$ set to zero and the shape for the indicated
  positive value of the corresponding coefficient, respectively. The blue dashed
  curves in the $\alpha_2$ and $\alpha_4$ panels show the shapes for the
  corresponding negative values.}
  \label{fig:cassini_alphan}
\end{figure}

The collective coordinates used in the present calculation are
$\{\alpha,\alpha_1,\alpha_2,\alpha_3,\alpha_4,\alpha_5\}$.
Figure~\ref{fig:cassini_alphan} illustrates how each of the five expansion coefficients
$\alpha_n$ changes the nuclear shape at a fixed elongation of $\alpha=0.98$. The odd
coefficients $\alpha_1$, $\alpha_3$, and $\alpha_5$ describe asymmetric deformations.
Here, $\alpha_1$ is mainly associated with mass asymmetry, $\alpha_3$ controls the
difference in prolate deformation between the two nascent fragments, and $\alpha_5$
controls the difference in higher-order shape distortion between the two nascent
fragments. The even coefficients $\alpha_2$ and $\alpha_4$ describe symmetric
deformations. For the even coefficients, $\alpha_2$ mainly changes the neck region,
whereas $\alpha_4$ changes the overall elongation and the distance between the fragment
centers.

\subsection{Multidimensional Langevin equation}

Fission dynamics is calculated by solving the Langevin equations in the
six-dimensional Cassini coordinate space. The deformation parameters
$\{\alpha,\alpha_1,\alpha_2,\alpha_3,\alpha_4,\alpha_5\}$ are denoted by
$\{q_i\}$. The equations
of motion for $\{q_i\}$ and their conjugate momenta $\{p_i\}$ are written as
\begin{align}
  \frac{dq_i}{dt}
  &= m_{ij}^{-1}p_j,
  \nonumber\\
  \frac{dp_i}{dt}
  &=
  -\gamma_{ij}m_{jk}^{-1}p_k
  -\frac{\partial F}{\partial q_i}
  -\frac{1}{2}\frac{\partial m_{jk}^{-1}}{\partial q_i}p_jp_k
  +g_{ij}R_j,
  \label{eq:Lan1}
\end{align}
where $m_{ij}$ and $\gamma_{ij}$ are the collective inertia and friction
tensors, respectively, $F$ is the free energy for the collective motion, and
$g_{ij}$ is the random-force strength tensor. The stochastic force $R_j$ is a
white-noise term with $\langle R_i(t)\rangle=0$ and
$\langle R_i(t)R_j(t')\rangle=2\delta_{ij}\delta(t-t')$, where
$\langle\cdots\rangle$ denotes an ensemble average. The strength tensor
$g_{ij}$ is determined from the modified Einstein relation
$T^*\gamma_{ij}=g_{ik}g_{jk}$, where $T^*$ is the effective nuclear
temperature calculated using the prescription described in
Refs.~\cite{hofmann1979,usang2017}.

The free energy is calculated in a macroscopic--microscopic framework
\cite{moller2016} as
\begin{align}
  F(\bm{q},T)
  &= F_{\mathrm{FRLDM}}(\bm{q},T)+F_{\mathrm{micro}}(\bm{q},T),
  \nonumber\\
  F_{\mathrm{FRLDM}}(\bm{q},T)
  &= V_{\mathrm{FRLDM}}(\bm{q})-a(\bm{q})T^2,
  \label{eq:free_energy}
\end{align}
where $F_{\mathrm{FRLDM}}$ and $F_{\mathrm{micro}}$ are the finite-range
liquid-drop model (FRLDM) term and the microscopic term in the free energy,
respectively. The
quantity $V_{\mathrm{FRLDM}}$ is the FRLDM potential, and $a$ is the level
density parameter \cite{toke1981}. The microscopic part
$F_{\mathrm{micro}}$ is evaluated with the Strutinsky shell correction method
\cite{strutinsky1967,strutinsky1968}, using single-particle levels obtained
from a deformed Woods-Saxon potential. Finite-temperature occupation numbers
are included in the evaluation of $F_{\mathrm{micro}}$ to account for the
damping of shell effects with increasing excitation energy
\cite{ivanyuk1999,ivanyuk2018}. The nuclear temperature $T$ is determined at
each time step from the relation $E^*=aT^2$. The excitation energy $E^*$ is
calculated from conservation of total energy as
\begin{align}
  E^*
  &=
  E_{\mathrm{full}}
  -\frac{1}{2}m_{ij}^{-1}p_ip_j
  -F(\bm{q},T=0),
  \label{eq:excitation}
\end{align}
where $E_{\mathrm{full}}$ is the total energy of the system.

The collective inertia and friction tensors are calculated microscopically
within linear response theory for nuclear collective motion
\cite{ivanyuk1999}. At finite temperature, the friction tensor is given by
\begin{align}
  \gamma_{ij}
  &=
  2\hbar \sum_{\mu\nu}
  (n_\mu^T-n_\nu^T)\xi_{\nu\mu}^2
  \frac{E_{\nu\mu}^- \Gamma_{\nu\mu}}
  {\left[(E_{\nu\mu}^-)^2+\Gamma_{\nu\mu}^2\right]^2}
  \nonumber\\
  &\qquad \times F_{\nu\mu}^iF_{\mu\nu}^j
  \nonumber\\
  &\quad
  +
  2\hbar \sum_{\mu\nu}
  (1-n_\mu^T-n_\nu^T)\eta_{\nu\mu}^2
  \frac{E_{\nu\mu}^+ \Gamma_{\nu\mu}}
  {\left[(E_{\nu\mu}^+)^2+\Gamma_{\nu\mu}^2\right]^2}
  \nonumber\\
  &\qquad \times F_{\nu\mu}^iF_{\mu\nu}^j,
  \label{eq:gamma_micro}
\end{align}
and the corresponding inertia tensor is given by
\begin{align}
  m_{ij}
  &=
  \hbar^2 \sum_{\mu\nu}
  (n_\mu^T-n_\nu^T)\xi_{\nu\mu}^2
  \frac{E_{\nu\mu}^- \left[(E_{\nu\mu}^-)^2-3\Gamma_{\nu\mu}^2\right]}
  {\left[(E_{\nu\mu}^-)^2+\Gamma_{\nu\mu}^2\right]^{3}}
  \nonumber\\
  &\qquad \times F_{\nu\mu}^iF_{\mu\nu}^j
  \nonumber\\
  &\quad
  +
  \hbar^2 \sum_{\mu\nu}
  (1-n_\mu^T-n_\nu^T)\eta_{\nu\mu}^2
  \frac{E_{\nu\mu}^+ \left[(E_{\nu\mu}^+)^2-3\Gamma_{\nu\mu}^2\right]}
  {\left[(E_{\nu\mu}^+)^2+\Gamma_{\nu\mu}^2\right]^{3}}
  \nonumber\\
  &\qquad \times F_{\nu\mu}^iF_{\mu\nu}^j.
  \label{eq:m_micro}
\end{align}
Here $\mu$ and $\nu$ label quasiparticle states, $E_\nu$ is the quasiparticle
energy, and $n_\nu^T=[1+\exp(E_\nu/T)]^{-1}$ is the
Fermi-Dirac occupation
factor. The quantity $\Gamma_{\nu\mu}$ is the collisional damping width for the
response involving quasiparticle states $\nu$ and $\mu$. The
combinations $E_{\nu\mu}^{\pm}=E_\nu\pm E_\mu$,
$\eta_{\nu\mu}=u_\nu v_\mu + u_\mu v_\nu$, and
$\xi_{\nu\mu}=u_\nu u_\mu - v_\mu v_\nu$ are defined in terms of the
Bogoliubov-Valatin transformation coefficients $u_\nu$ and $v_\nu$. The matrix
element $F_{\mu\nu}^i$ depends on the mean-field Hamiltonian; through this
dependence, shell structure enters the transport coefficients. These
microscopic expressions depend explicitly on $T$. In
general, the friction tensor increases with temperature, whereas the inertia
tensor decreases.

The Langevin trajectories are started from the ground-state configuration of
$^{236}$U. A
scission configuration is identified when the neck radius becomes
$2~\mathrm{fm}$ or smaller. A total of $1 \times 10^6$ Langevin trajectories
are generated, and the calculation is continued until
$5 \times 10^5$ trajectories have reached scission.

\begin{figure}
  \includegraphics[width=\columnwidth, trim=0.30cm 3.90cm 1.42cm 7.55cm, clip]{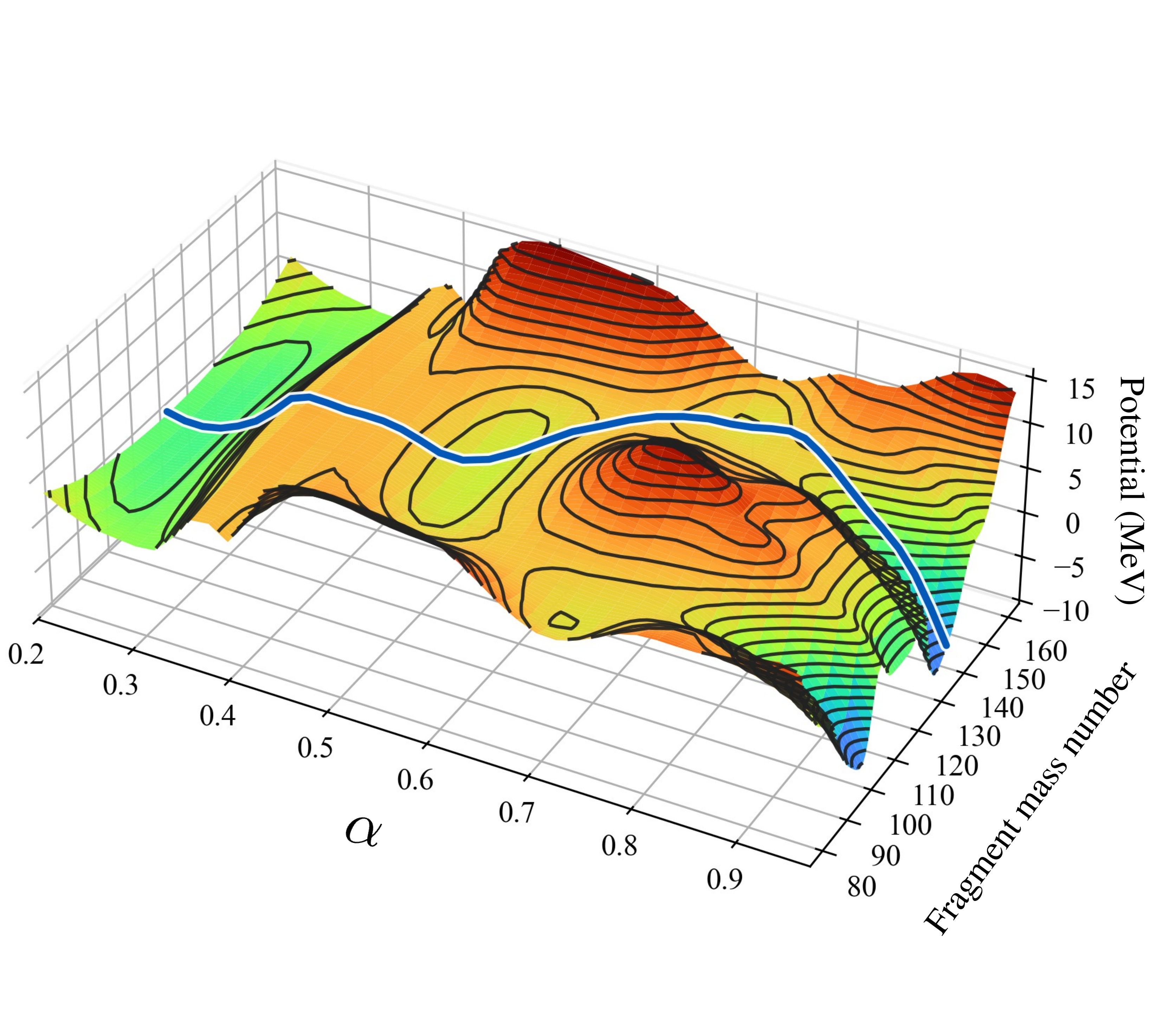}
  \caption{Potential energy surface on the $(\alpha,A_F)$ plane reconstructed
  from the ensemble of Langevin trajectories. The blue curve traces the peak positions of
  the trajectory distribution projected onto the same plane. The potential energy is
  measured from the ground-state energy of $^{236}$U, and the contour
  interval is 1~MeV.}
  \label{fig:pot}
\end{figure}

Figure~\ref{fig:pot} shows a potential energy surface (PES) reconstructed on the
$(\alpha,A_F)$ plane for $^{236}$U. Here, $A_F$ denotes the fragment mass number defined
at scission. For the PES visualization, this definition is extended to pre-scission
configurations, and $A_F$ is used as a mass-asymmetry coordinate. The potential is
defined in the six-dimensional Cassini coordinate space; therefore, the remaining
Cassini parameters must be specified to visualize it on the $(\alpha,A_F)$ plane. To
construct this projected surface, the Cassini parameter sets recorded along the
trajectories are grouped into bins on the $(\alpha,A_F)$ mesh. In each populated bin,
the last parameter set recorded in that bin is retained for each Langevin event that
visits it. These parameter sets are averaged over the events to define a representative
shape, for which the potential is calculated. The representative path shown in
Fig.~\ref{fig:pot} is obtained by
tracing the peak positions of the trajectory distribution projected onto the same plane.

The representative path remains in the mass-symmetric region up to the second minimum.
At and beyond the second fission barrier, the representative path moves toward
mass-asymmetric shapes and follows the main asymmetric valley. The projected PES alone
does not show distinct signatures in mass asymmetry for the Standard~I and Standard~II
modes. However, a second low-energy region associated with the SL component is visible
near symmetric mass division in the reconstructed PES.

At the scission point, the nuclear shape is divided into two fragment regions.
For each fragment $F$, we define the mass multipole moment and the corresponding
deformation parameter as
\begin{align}
  Q_{l0}^{(F)}
  &=
  \int_{V_F}\rho(\bm{r})\,r^lY_{l0}(\theta)\,d^3r,
  \nonumber\\
  \beta_{l,F}
  &=
  \frac{4\pi}{3A_FR_F^l}Q_{l0}^{(F)}.
  \label{eq:beta_def}
\end{align}
Here $l$ is the multipole order, $Y_{l0}$ is a spherical harmonic, and $V_F$
is the part of the nuclear volume assigned to fragment $F$ by the neck plane.
The quantity $R_F$ is the radius of a sphere with the same volume
as $V_F$. The nucleon density $\rho(\bm{r})$ is taken to be uniform within the
volume enclosed by the nuclear surface at scission and zero outside. The
coordinates $(r,\theta)$
are defined with the origin at the center of mass of fragment $F$.

The TKE is evaluated at scission as the sum of the Coulomb repulsion energy between the
two nascent fragments and their prescission kinetic energy. The Coulomb contribution is
dominated by the distance between the fragment charge centers at the scission
configuration.

\subsection{Fission mode classification}\label{sec:kmeans}

The scission configuration of each Langevin event is obtained in the six-dimensional
Cassini coordinate space. However, direct analysis of the six Cassini coordinates is not
well suited for mode classification because fragment mass and deformations depend on
combinations of the coefficients. For this reason, the classification is based on
scission quantities constructed from the resulting shape. In this work, we use the
fragment mass and the quadrupole deformations of both fragments,
$(A_F,\beta_{2,F},\beta_{2,\bar F})$, for mode classification. The symbol $\bar F$
denotes the partner of $F$. TKE is strongly related to the fragment deformations and is
therefore excluded from the classification variables. The octupole deformation $\beta_3$
is also excluded to minimize the number of variables required for classification. Both
TKE and $\beta_3$ are examined only after the mode assignment.

An unsupervised classification is performed using the $k$-means algorithm
\cite{macqueen1967,lloyd1982}. On the basis of the phenomenology of actinide fission,
the classification results are interpreted in terms of three physical modes: Asymmetric
1 (AS1), Asymmetric 2 (AS2), and SL. The mass and TKE characteristics of the modes
identified by the present classification do not necessarily coincide with those of the
conventional Standard~I and Standard~II modes. Therefore, the labels AS1 and AS2 are
used to distinguish the present modes from the conventional ones. Details of the mode
classification are given in Appendix~\ref{app:kmeans}.

\section{Results and discussion}\label{sec:results}

\subsection{Decomposition of fission modes}\label{sec:massdist}

\begin{figure}
  \includegraphics[width=\columnwidth, trim=1.62cm 1.78cm 6.17cm 2.25cm, clip]{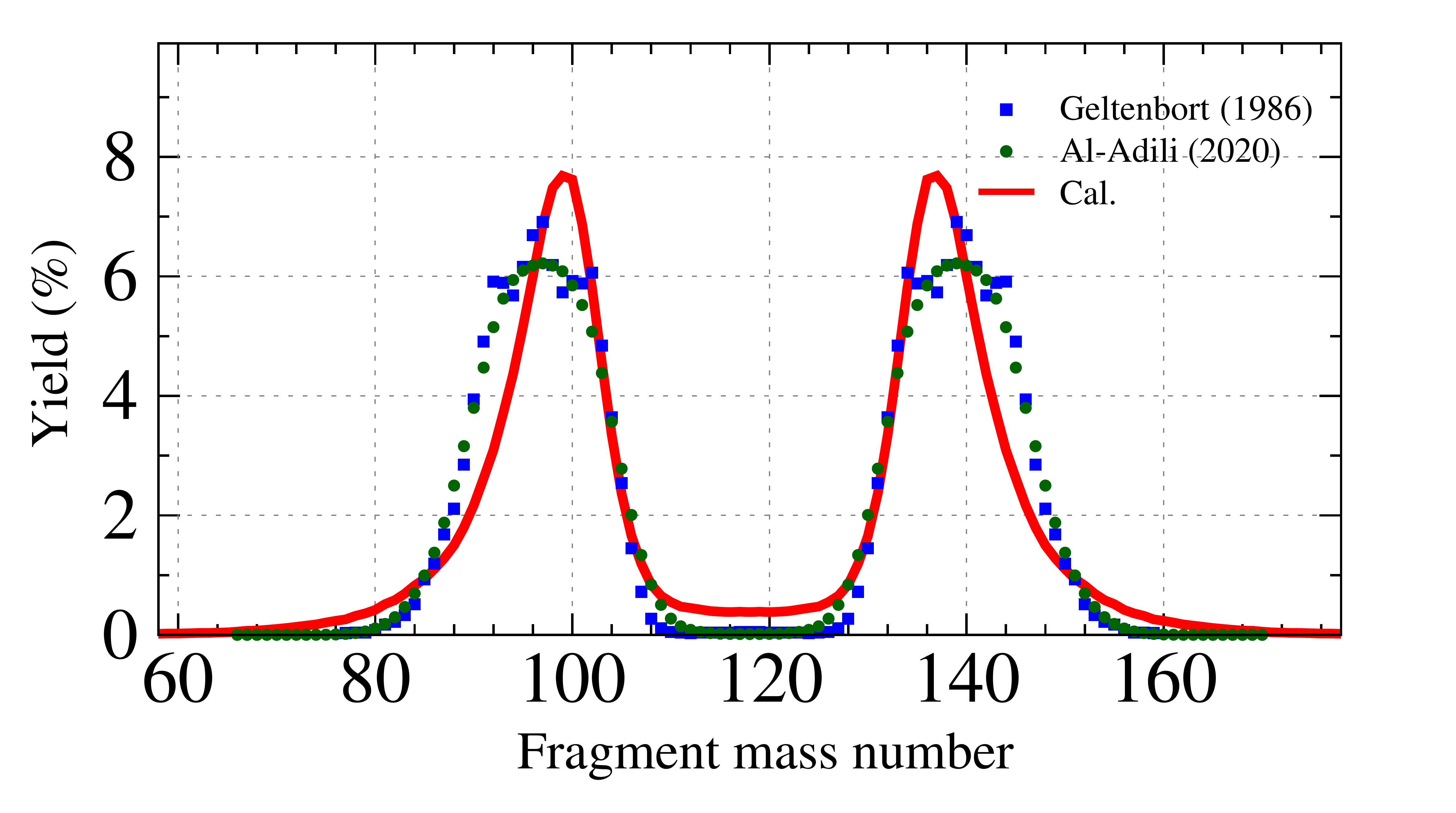}
  \caption{Pre-neutron fragment mass distribution for thermal neutron-induced fission
  of $^{235}$U. The red line shows the calculated distribution. The blue squares
  and green circles denote the experimental data given in
  \cite{geltenbort1986} and \cite{aladili2020}, respectively.}
  \label{fig:FMD}
\end{figure}

Figure~\ref{fig:FMD} shows the pre-neutron fragment mass distribution obtained
from the six-dimensional Langevin calculation for thermal neutron-induced
fission of $^{235}$U. The calculated distribution approximately reproduces the
dominant asymmetric peaks in the experimental
data~\cite{geltenbort1986,aladili2020}, although the yield around symmetric mass
division is slightly overestimated.

\begin{figure}[t]
  \centering
  \includegraphics[width=\columnwidth, trim=1.32cm 1.42cm 7.02cm 1.95cm, clip]{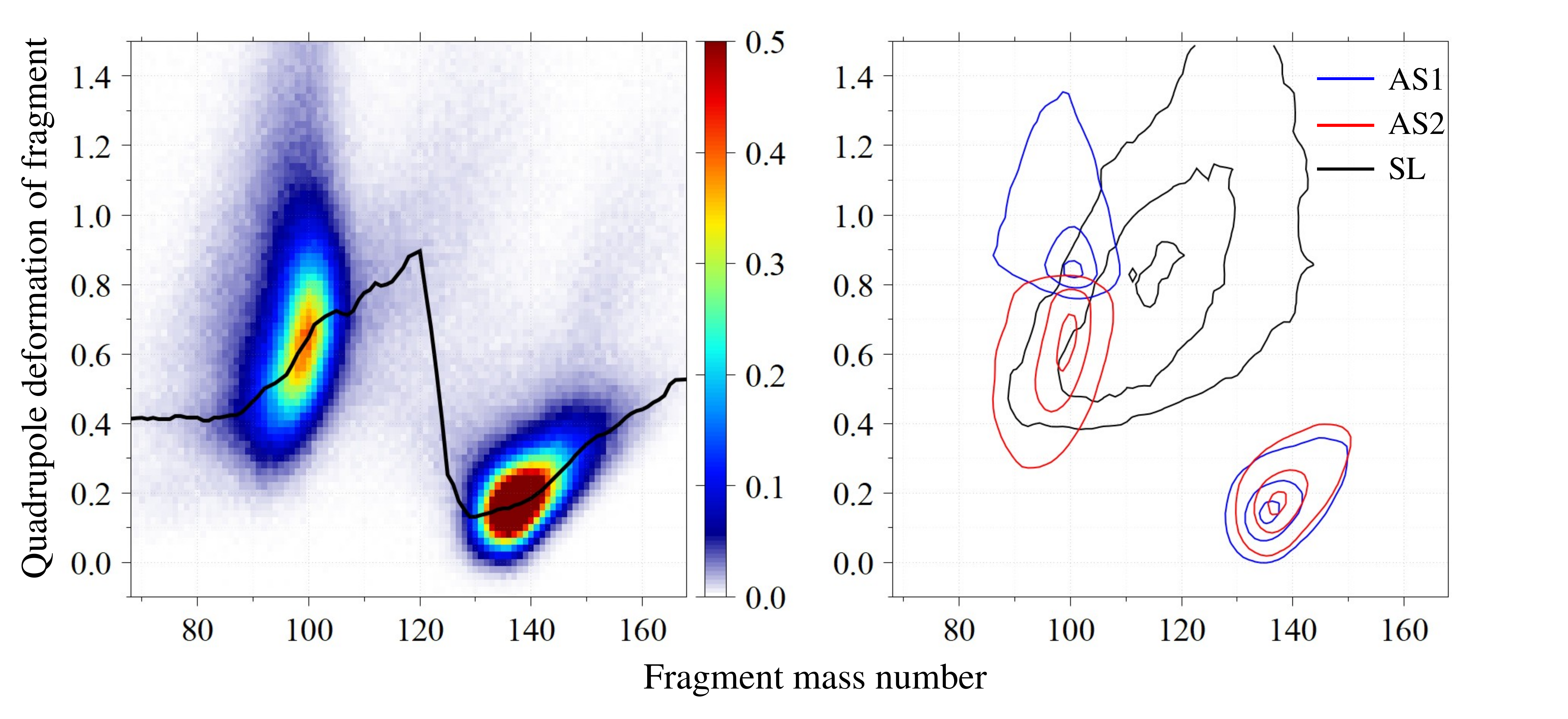}
  \caption{
  Distribution of fragment mass number $A_F$ and quadrupole deformation $\beta_{2,F}$ at scission
  for thermal neutron-induced fission of $^{235}$U. The distribution before mode classification
  is shown in the left panel, with the peak of the $\beta_{2,F}$ distribution at each $A_F$
  indicated by the black curve. The AS1 (blue), AS2 (red), and SL (black) contours after mode assignment
  are shown in the right panel. For each mode, the contours are drawn
  at 10\%, 50\%, and 90\% of the maximum projected density on the $(A_F,\beta_{2,F})$ plane.
  }
  \label{fig:b2fmd}
\end{figure}

Figure~\ref{fig:b2fmd} shows the distribution of fragment mass number $A_F$ and quadrupole
deformation $\beta_{2,F}$ at scission. The left panel shows the distribution before the
mode classification. On the heavy fragment side ($130\lesssim A_H\lesssim145$), the
distribution is concentrated within a narrow range of small $\beta_{2,F}$ values. On the
light fragment side near $A_L\approx100$, the distribution extends from moderately
deformed to strongly elongated shapes. A broad distribution of the quadrupole
deformation of the light fragment has also been reported in four-dimensional Langevin
calculations based on the TCSM \cite{ishizuka2020,shimada2021}.

The events are classified using fragment mass and the quadrupole deformations
of both fragments, as described in Sec.~\ref{sec:kmeans}. The right panel of
Fig.~\ref{fig:b2fmd} shows the resulting mode distributions projected onto the
$(A_F,\beta_{2,F})$ plane.
For each mode, the contours are drawn at 10\%, 50\%, and 90\% of the
maximum projected density.

The contours show that AS1 and AS2 are separated on the light-fragment side but largely
overlap on the heavy-fragment side. AS2 has moderately deformed light fragments, whereas
AS1 has more elongated ones. The SL component is distributed in the symmetric mass
region, where both fragments have large $\beta_2$ values.

Based on the classification shown in Fig.~\ref{fig:b2fmd},
Table~\ref{tab:Umode_mode_average} summarizes the representative scission
shape, fragment masses and deformation parameters, mode-averaged TKE, and yield
for each mode. Each representative shape is constructed from the mode-averaged
Cassini parameters at scission.

The two asymmetric modes exhibit distinct mass splits and fragment deformation patterns.
AS1 corresponds to $(A_L,A_H)=(99,137)$ and combines a strongly elongated light
fragment, $(\beta_{2,L},\beta_{3,L})=(1.03,0.51)$, with a relatively compact heavy
fragment, $(\beta_{2,H},\beta_{3,H})=(0.20,0.14)$. AS2 corresponds to
$(A_L,A_H)=(96,140)$ and combines a moderately deformed light fragment,
$(\beta_{2,L},\beta_{3,L})=(0.58,0.33)$, with a heavy fragment characterized by
$(\beta_{2,H},\beta_{3,H})=(0.23,0.19)$. The difference between the two modes is more
pronounced in the light fragment deformation. The representative heavy fragment masses
of AS1 and AS2 are close to
those of Standard~I and Standard~II, respectively, in the three-mode description of
$^{235}$U(n, f)~\cite{knitter1987,brosa1990}. In contrast to the relative TKE values of
Standard~I and Standard~II, AS2 has a mode-averaged TKE of 177.3~MeV, higher than the AS1
value of 164.4~MeV. The lower TKE of AS1 is associated with the larger elongation of its
light fragment.

The mass asymmetries of AS1 and AS2, identified from the calculated
$(A_F,\beta_{2,F},\beta_{2,\bar F})$ distribution for $^{236}$U, are consistent with the
mass asymmetries of Standard~I and Standard~II, respectively, inferred from the
experimental mass--TKE distribution. Applying the same analysis to $^{258}$Md identifies two
asymmetric modes whose mass asymmetries differ substantially from those in
$^{236}$U~\cite{okada2026md}. The conventional Standard~I/Standard~II description leaves
such nucleus-dependent changes unspecified.

The SL mode corresponds to the symmetric split $A_L=A_H=118$. Both fragments
are elongated, with identical deformation parameters
$(\beta_{2,L},\beta_{3,L})=(\beta_{2,H},\beta_{3,H})=(0.88,0.22)$. It has the
lowest mode-averaged TKE of 150.3~MeV.

The results show that AS2 gives the dominant contribution, 68.6\%, whereas
AS1 makes a smaller asymmetric contribution, 27.1\%, and SL remains a minor
symmetric component, 4.3\%.

The classification is examined for fission from a more highly excited state using
scission events
calculated for 14~MeV neutron-induced fission of $^{235}$U. The results are summarized
in Table~\ref{tab:n14_mode_average} in Appendix~\ref{app:n14}. The characteristic fragment
deformation patterns are preserved at the higher excitation energy: AS1 combines a
strongly elongated light fragment with a compact heavy fragment, AS2 combines a
moderately deformed light fragment with a similarly compact heavy fragment, and SL has
two significantly elongated fragments. Compared with the case of thermal neutron-induced fission, the AS1 yield
decreases and the SL yield increases, whereas the AS2 yield changes only slightly.

\newcommand{\modeshapegraphic}[1]{%
  \raisebox{0.5ex}{\includegraphics[width=0.18\columnwidth, trim=1.2cm 6.0cm 1.2cm 6.0cm, clip]{#1}}%
}

\begin{table}[t]
  \centering
  \caption{Representative scission configurations and associated physical quantities
  for thermal neutron-induced fission of $^{235}$U. For each mode, the fragment
  mass numbers are indicated inside the representative shape, and the
  quadrupole and octupole deformations of each fragment, TKE, and yield are
  listed.}
  \label{tab:Umode_mode_average}
  \setlength{\tabcolsep}{4pt}
  \begin{tabular}{ l|ccc }
    \hline
    Mode & AS1 & AS2 & SL \\
    \hline
    \raisebox{2.5ex}{Shape} \rule{0pt}{6.5ex}
    & \modeshapegraphic{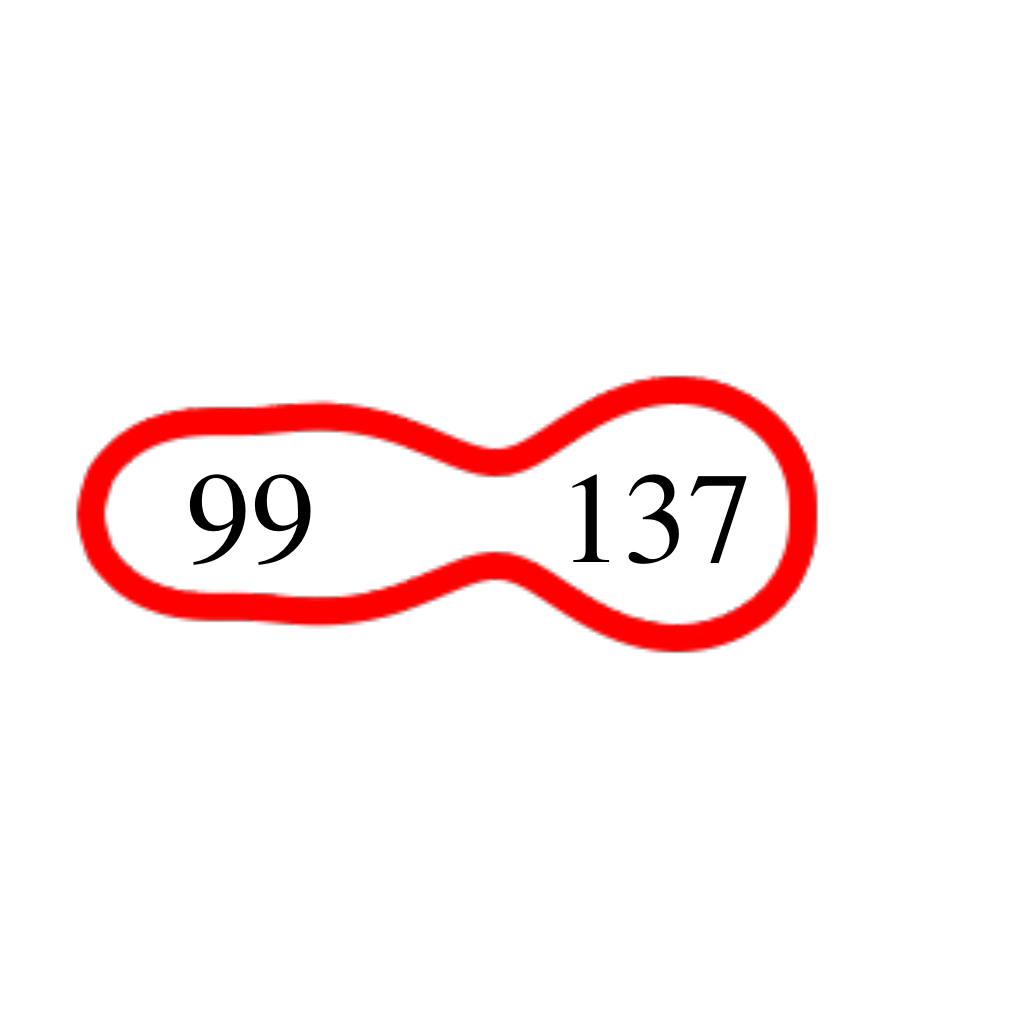}
    & \modeshapegraphic{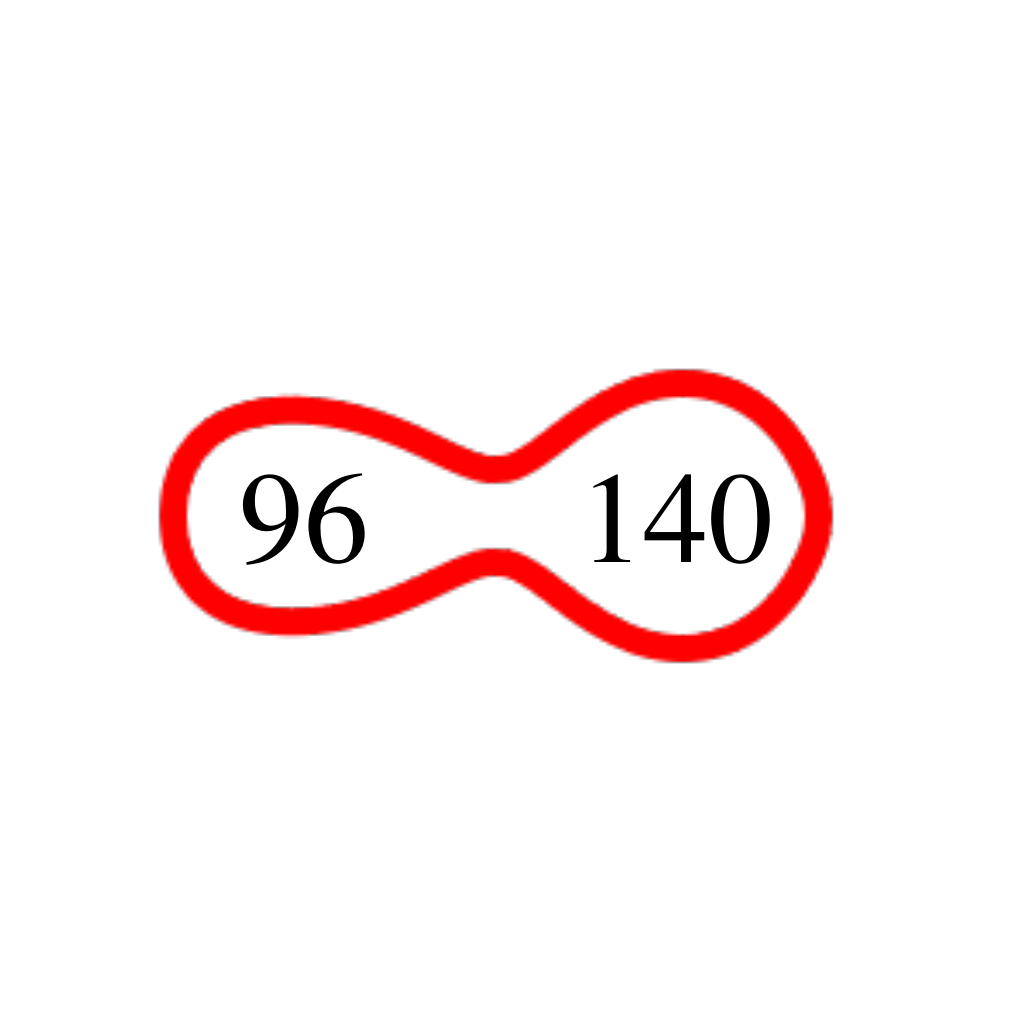}
    & \modeshapegraphic{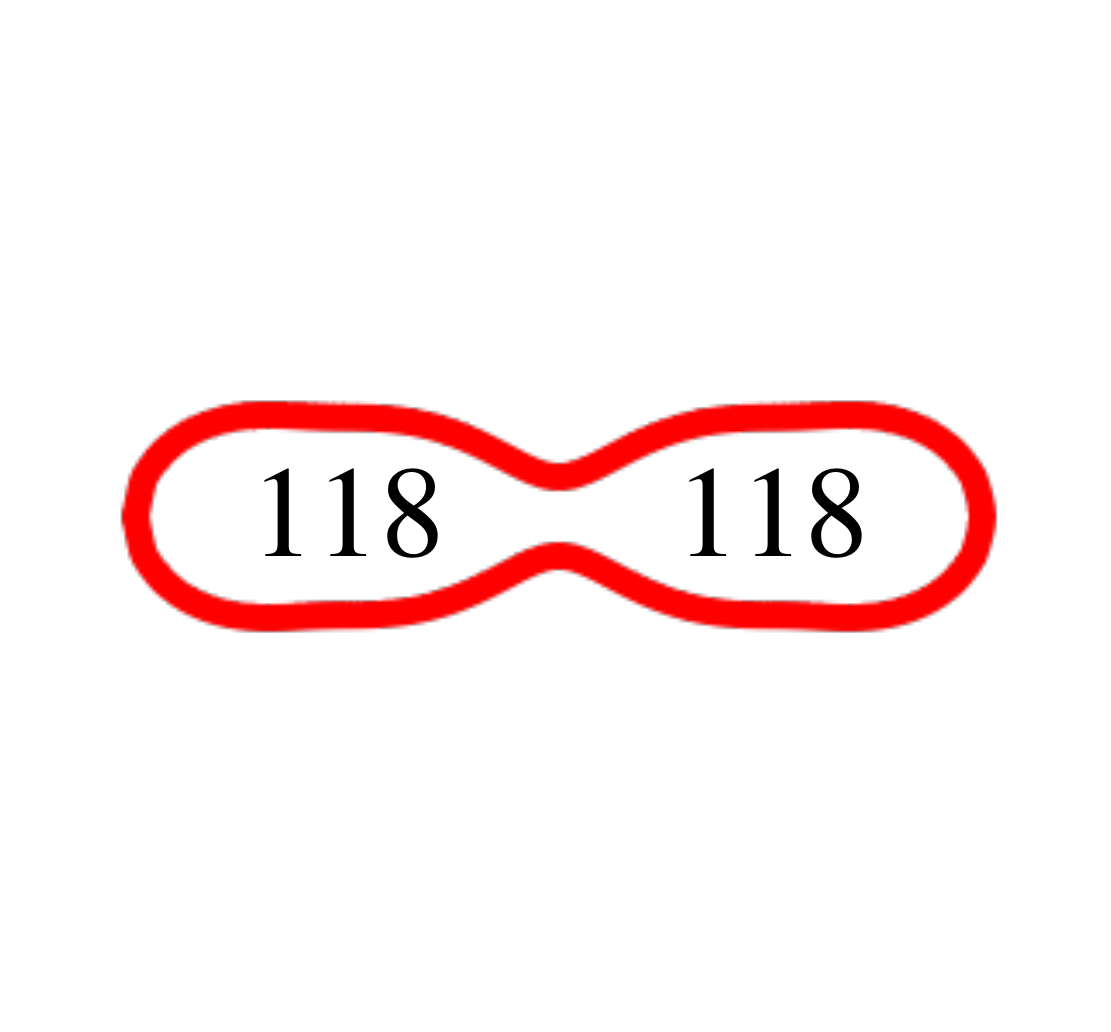} \\
    \hline
    $\beta_{2,L}$, $\beta_{2,H}$ & 1.03, 0.20 & 0.58, 0.23 & 0.88, 0.88 \\
    $\beta_{3,L}$, $\beta_{3,H}$ & 0.51, 0.14 & 0.33, 0.19 & 0.22, 0.22 \\
    TKE (MeV)            & 164.4      & 177.3      & 150.3 \\
    Yield (\%)           & 27.1       & 68.6       & 4.3 \\
    \hline
  \end{tabular}
\end{table}

\subsection{Fission-mode dependence of total kinetic energy}\label{sec:tke}

\begin{figure}
  \includegraphics[width=\columnwidth, trim=1.30cm 1.35cm 0.50cm 5.30cm, clip]{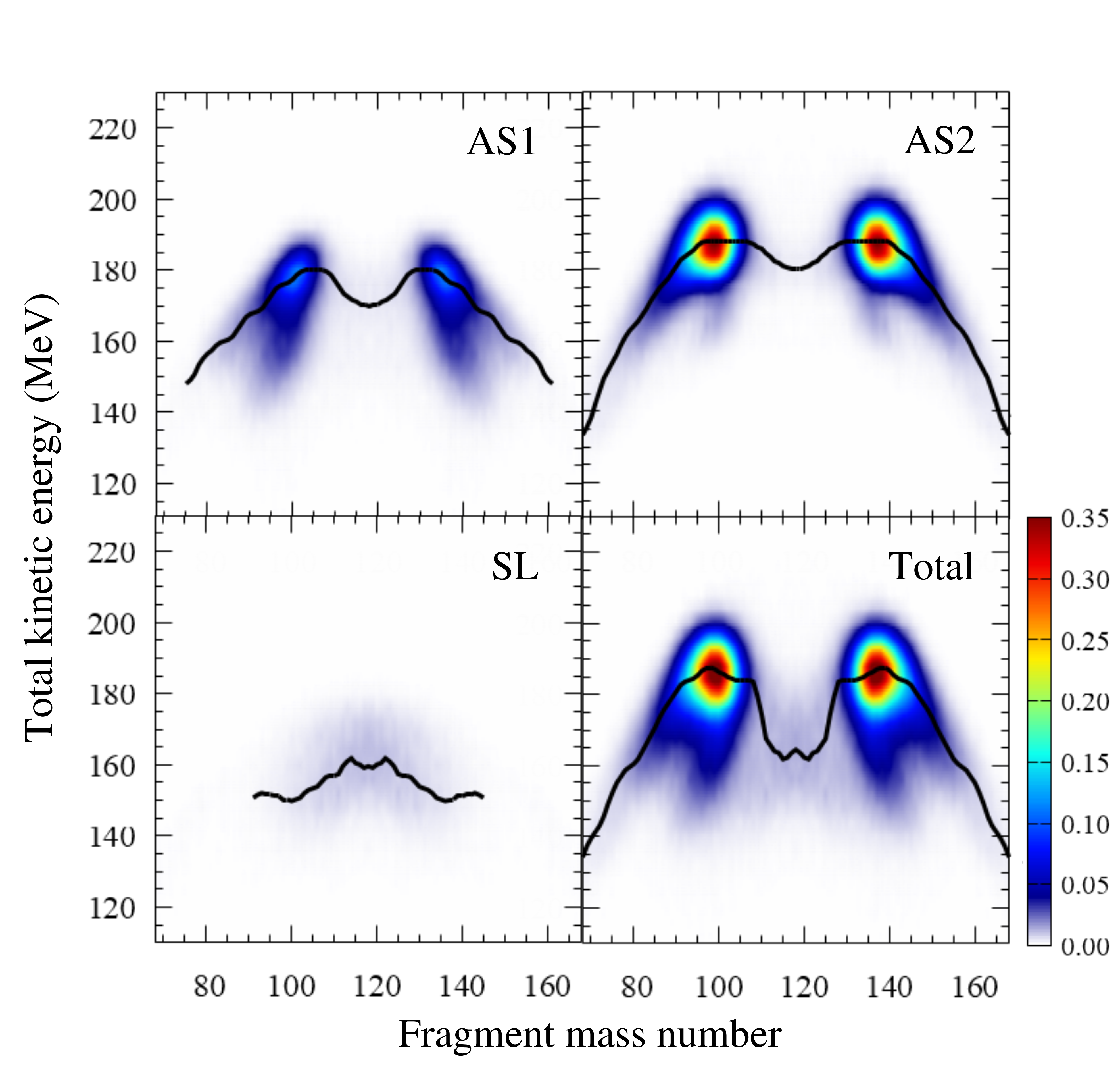}
  \caption{Mass--TKE distributions for thermal neutron-induced fission of $^{235}$U.
  The upper-left, upper-right, lower-left, and lower-right panels show the AS1, AS2,
  SL, and total distributions, respectively. The color scale represents the density
  in bins with widths $\Delta A_F=1$ and $\Delta \mathrm{TKE}=4~\mathrm{MeV}$.
  The total distribution is normalized to 200\%.
  The black curve in each panel indicates the TKE corresponding to the peak of the distribution at each fragment mass number.}
  \label{fig:TKEFMD_nth}
\end{figure}

The scission events assigned to each mode in the classification shown in
Fig.~\ref{fig:b2fmd} are projected onto the fragment mass--TKE plane while
retaining their mode labels. Figure~\ref{fig:TKEFMD_nth} shows the resulting
AS1, AS2, and SL distributions and the total distribution for thermal
neutron-induced fission of $^{235}$U. In the asymmetric mass region, the AS1
distribution extends toward lower TKE, whereas the AS2 distribution occupies a
broader asymmetric mass range and lies mainly at higher TKE. The SL
distribution is concentrated around symmetric mass division and at low TKE.

\begin{figure}
  \includegraphics[width=\columnwidth, trim=1.45cm 1.55cm 0.20cm 1.75cm, clip]{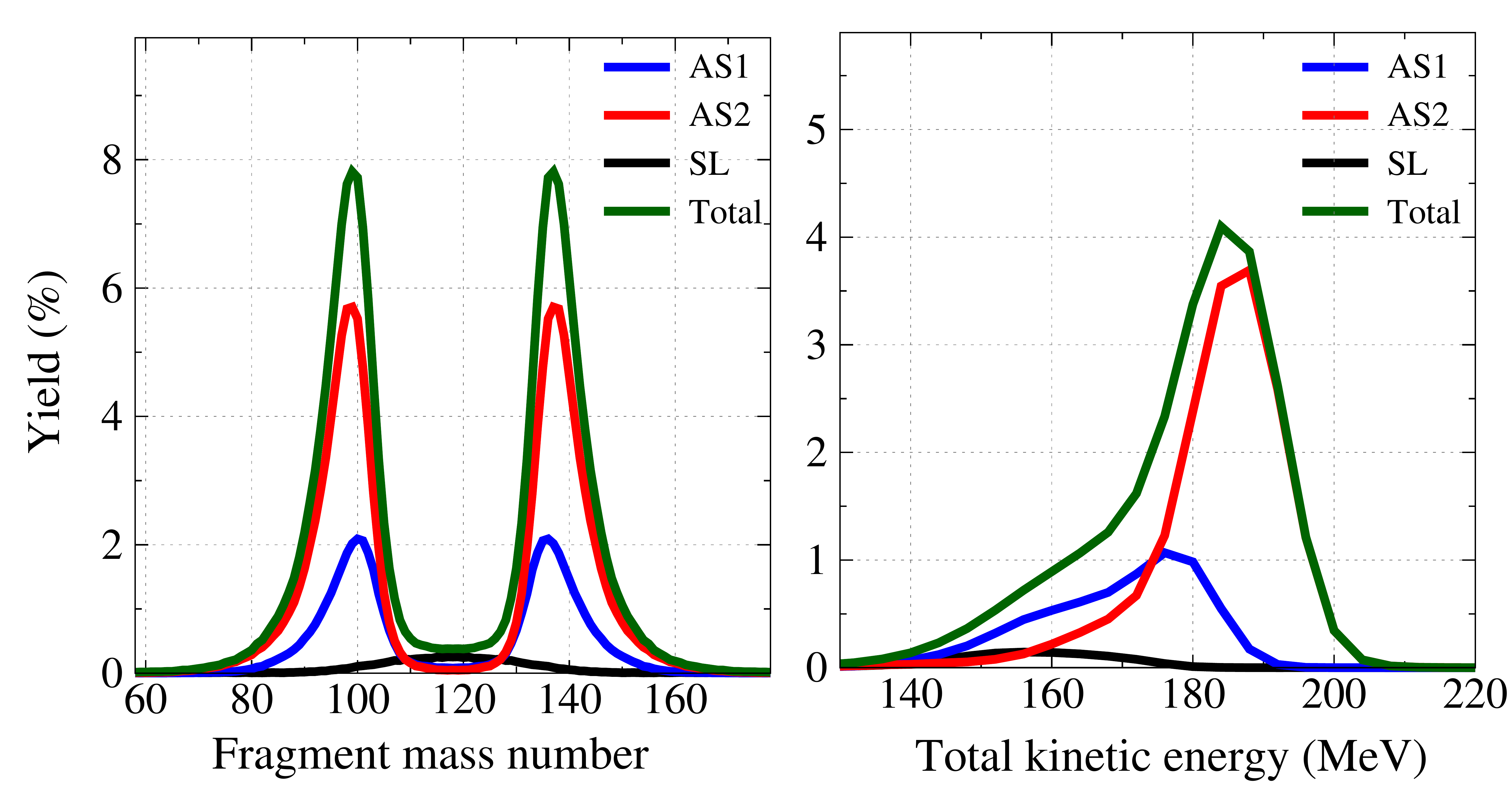}
  \caption{One-dimensional projections of the mass--TKE distributions for each mode
  in thermal neutron-induced fission of $^{235}$U. The left and right panels show
  the fragment mass and TKE distributions, respectively.}
  \label{fig:mode}
\end{figure}

Figure~\ref{fig:mode} shows the one-dimensional fragment mass and TKE distributions
obtained by projecting the AS1, AS2, and SL mass--TKE distributions in
Fig.~\ref{fig:TKEFMD_nth} onto the respective axes. In the fragment mass distribution,
AS1 and AS2 overlap in the asymmetric mass region; AS2 gives the dominant contribution
there, whereas the AS1 distribution is shifted toward smaller mass asymmetry. SL gives
the dominant contribution in the symmetric mass region. In the TKE distribution, AS2
dominates at high TKE, whereas AS1 and SL contribute at lower TKE. Overall, the three
modes overlap extensively in both the fragment mass and TKE distributions.

\begin{figure}
  \includegraphics[width=\columnwidth, trim=1.00cm 1.10cm 1.90cm 0.70cm, clip]{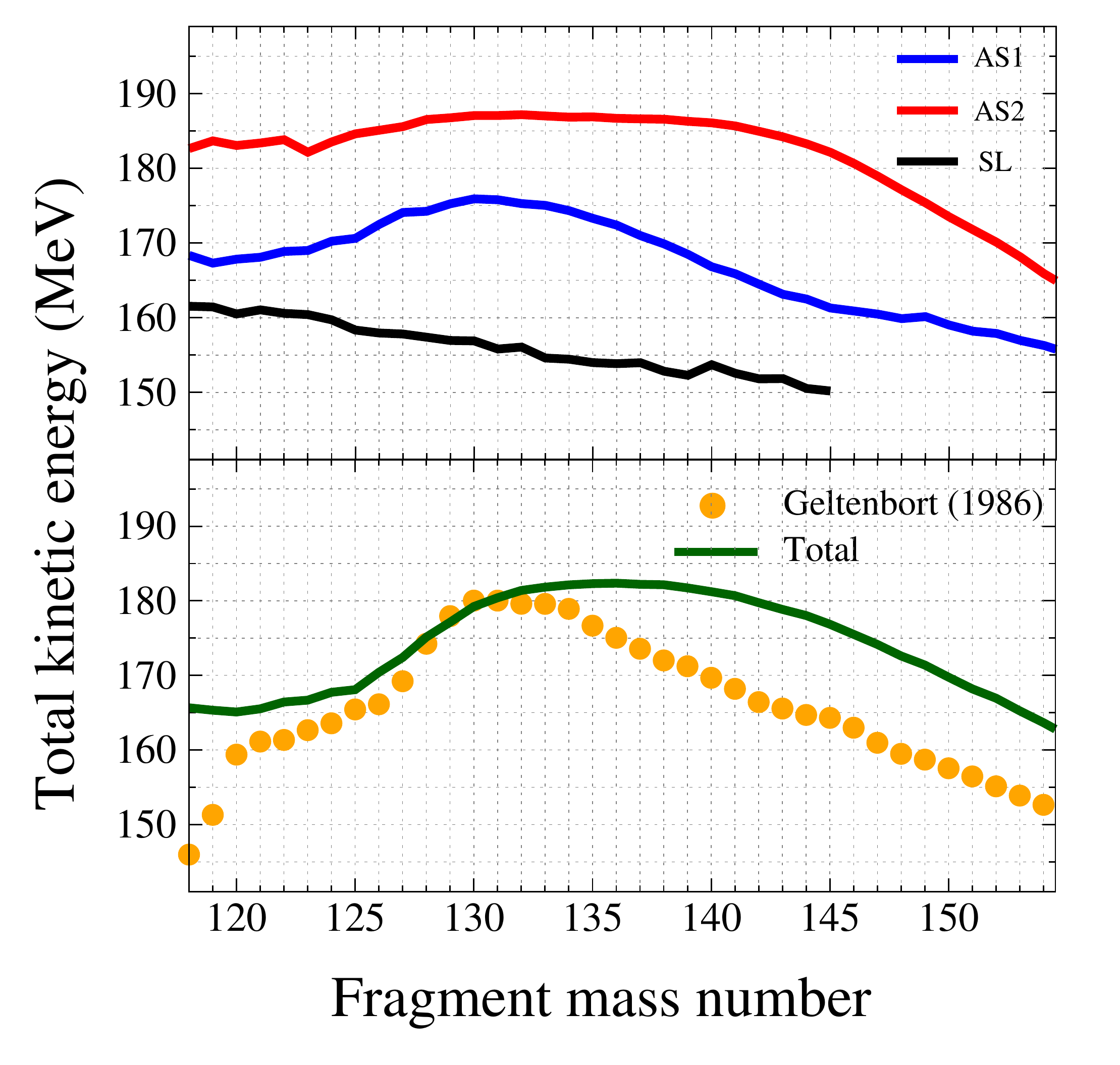}
  \caption{Average TKE as a function of pre-neutron fragment mass for thermal neutron-induced
  fission of $^{235}$U. In the upper panel, the blue, red, and black curves show
  the average TKE for AS1, AS2, and SL, respectively. In the lower panel, the
  green curve shows the total average TKE, and the orange circles denote the
  experimental data given in \cite{geltenbort1986}.}
  \label{fig:aveTKE}
\end{figure}

The upper panel of Fig.~\ref{fig:aveTKE} shows the mass dependences of the average TKE
for the three modes. AS2 forms a plateau at high average TKE over a broad fragment mass
region and maintains higher average TKE values than the other modes. AS1 reaches a
maximum near $A_H\simeq130$ and then decreases with increasing heavy fragment mass,
whereas the average TKE of the SL mode shows an approximately monotonic decrease.

The lower panel of Fig.~\ref{fig:aveTKE} compares the total average TKE with the
experimental data from Ref.~\cite{geltenbort1986}. The calculated curve follows the
general trend of the data, rising from symmetric mass division to the
maximum
around $A_H=137$ and then decreasing toward larger heavy fragment masses. However, the
calculation overestimates the measured TKE near symmetric mass division and in the
asymmetric region with $A_H>135$.

\subsection{Single-particle levels of the representative fragments}
\label{sec:results_levels}

\begin{figure*}[t]
  \centering
  \includegraphics[width=\textwidth, trim=0.0cm 0.27cm 2.35cm 0.0cm, clip]{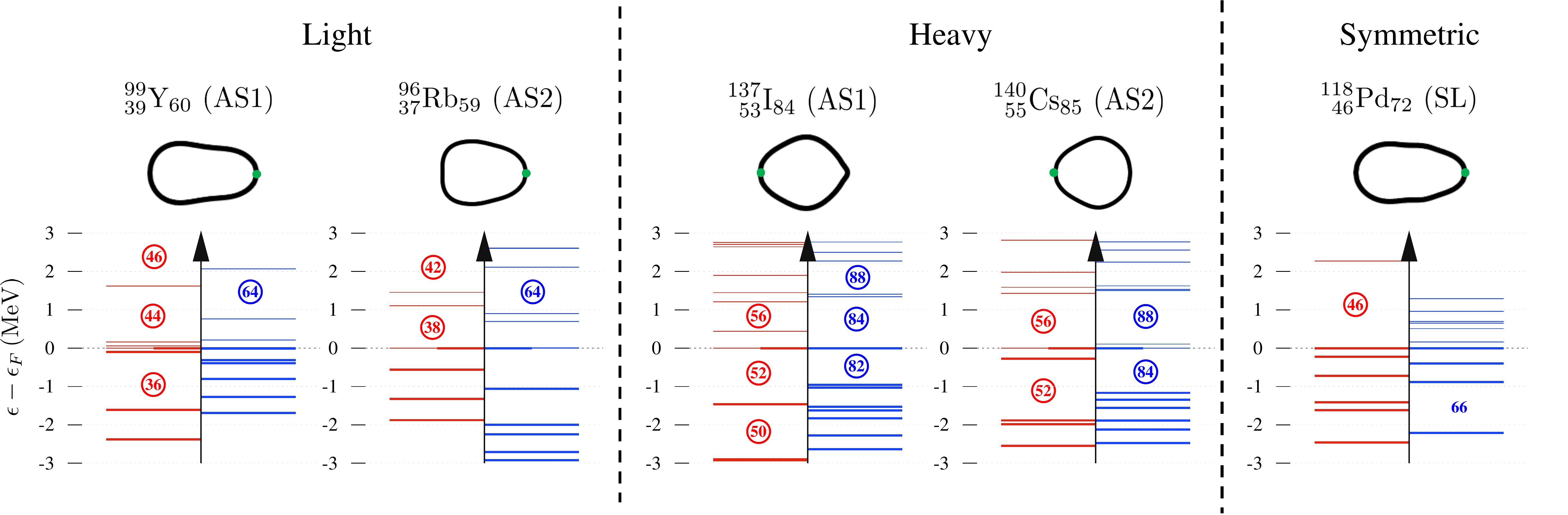}
  \caption{Proton and neutron single-particle levels of the representative
  fragments.
  From left to right, the panels show the AS1 and AS2 light fragments, the AS1 and AS2
  heavy fragments, and the symmetric SL fragment. Red and blue lines denote proton and neutron levels,
  respectively. Thick and thin lines denote occupied and unoccupied levels, respectively.
  For fragments with odd $Z$ or odd $N$, the corresponding level at the Fermi energy is drawn with both line
  widths. Energies are measured relative to the Fermi energy $\epsilon_F$ of each fragment.
  Circles mark the particle numbers corresponding to the level gaps discussed in the text.
  Green dots indicate the neck points of the fragment shapes.}
  \label{fig:levels}
\end{figure*}

The shell structures of the representative fragments are examined using proton
and neutron single-particle levels. For each representative scission
configuration shown in Table~\ref{tab:Umode_mode_average}, the shape of each
fragment is fitted separately with the Cassini shape parametrization for the
single-particle calculation. Integer values are then assigned to $Z$ and $N$,
with the $Z/N$ ratio kept close to that of $^{236}$U.

Figure~\ref{fig:levels} shows the resulting proton and neutron single-particle
levels. The isotope label for each fragment corresponds to a representative
point in $(Z,N)$ space obtained from the mode averages, whereas the shell
structure of the fragment is assessed from the level patterns near the Fermi
energy.

The middle block of Fig.~\ref{fig:levels} shows the single-particle levels of the AS1
and AS2 heavy fragments. For AS1, proton gaps appear at $Z=50$ and $Z=52$ and neutron
gaps at $N=82$ and $N=84$, indicating an intermediate shell structure with both
spherical and deformed features. The gaps at $Z=56$ and $N=88$ are only weakly
developed. This intermediate character is consistent with recent experimental analyses
associating Standard~I with the proton shell gap at
$Z=52$~\cite{martin2021,berriman2022,dey2025}. For AS2, the spherical shell gaps at
$Z=50$ and $N=82$ disappear, whereas well-developed gaps appear at $Z=56$ and $N=88$.
The AS2 shell structure is consistent with the shell features associated with quadrupole
and octupole deformations in the $^{144}$Ba region~\cite{scamps2018}. A deformed neutron
shell at $N=88$ was also identified in a scission-point model~\cite{wilkins1976}.

The left block of Fig.~\ref{fig:levels} shows the single-particle levels of the
representative light fragments for AS1 and AS2,
$^{99}_{39}\mathrm{Y}_{60}$ and $^{96}_{37}\mathrm{Rb}_{59}$, respectively. In
both modes, a neutron gap appears at $N=64$ near the Fermi energy, consistent
with a deformed neutron shell inferred from experimental fission fragment
systematics~\cite{veselsky2004,caamano2017,ramos2019}. The proton levels, by
contrast, show different gap structures for AS1 and AS2.

In the light fragment of AS2, proton gaps appear at $Z=38$ and $Z=42$. The
$Z=38$ gap also appears in a microscopic calculation of fragment
single-particle levels for $^{236}$U fission~\cite{bernard2023}. Figure~3 of
the cited study also shows an unlabeled level spacing corresponding to
$Z=42$, although this feature is not discussed in the text. The shell
features found in both AS2 fragments indicate that the representative scission
configuration obtained from the classified Langevin events reflects the
fragment shell structures predicted by the microscopic calculation for
$^{236}$U fission.

In the light fragment of AS1, proton gaps appear at $Z=36$, $Z=44$, and $Z=46$. Direct
correspondence between these gaps and proton shell gaps in the Standard~I light fragment
has been discussed only to a limited extent. However, similar proton numbers have been
associated with shell effects in strongly deformed fragments of lighter fissioning
systems. For mercury isotopes, quadrupole shell effects may contribute to the formation
of elongated heavy prefragments with $Z_H\simeq42$--46, as suggested by constrained
Hartree-Fock calculations with BCS pairing correlations~\cite{scamps2019}. Constrained
HFB calculations of $^{182}$Hg identify a proton shell gap at $Z=36$ in the strongly
elongated $^{82}$Kr light prefragment at $\beta_2\simeq1.4$~\cite{morfouace2025}. On the
experimental side, fragment charges near $Z=36$ and $Z=44$ were inferred from the
measured fragment mass split in $^{180}$Hg fission~\cite{andreyev2010}. Broader
systematics identify shell stabilization near $Z=36$~\cite{mahata2022} and shell-driven
structures in the $Z=34$--36 and $Z=44$--46 regions for fissioning systems below the
actinides~\cite{buete2025}. The similarity between the shell gaps in the AS1 light
fragment and those inferred experimentally and theoretically in lighter fissioning
systems suggests that related shell structures may occur across a broad range of
fissioning systems.

The right block of Fig.~\ref{fig:levels} shows the single-particle levels of the
representative SL fragment $^{118}_{46}\mathrm{Pd}_{72}$. A proton gap exceeding 2~MeV
appears at $Z=46$. A corresponding gap also appears in the AS1 light fragment but is
substantially larger in the SL proton spectrum. This pronounced gap suggests that proton
shell effects may contribute to the stability of the elongated SL configuration.
A neutron gap at $N=66$ is also visible below the Fermi energy, although no pronounced
neutron gap appears at the Fermi energy.

Overall, the single-particle level patterns reveal characteristic shell
structures in the representative fragments of all three modes, suggesting
that fragment shell effects contribute to the emergence of these fission
modes.

\section{Summary}\label{sec:summary}

Thermal neutron-induced fission of $^{235}$U was studied using a six-dimensional
Langevin calculation based on the Cassini shape parametrization. The scission events
were classified into three fission modes (AS1, AS2, and SL) by applying the $k$-means
algorithm to the fragment mass and the quadrupole deformations of both fragments.
Representative scission shapes were constructed, and pre-neutron fragment mass--TKE
distributions were obtained from the classified events for the three modes. Proton and
neutron single-particle levels were calculated for the representative fragments of each
mode to examine their shell structures. The calculated fragment mass distribution and
overall average TKE were compared with experimental data.

The calculated fragment mass distribution approximately reproduces the dominant
asymmetric peaks in the experimental data, although the symmetric yield is slightly
overestimated. AS2 provides the dominant contribution to the calculated yield. The
projection of the classified events onto the fragment mass--quadrupole deformation plane
shows characteristic distributions for the three modes, with AS1 and AS2 differing most
prominently in the quadrupole deformations of the light fragments. The average fragment
deformations show that AS1 combines a compact heavy fragment with a strongly elongated
light fragment, whereas AS2 combines a similarly compact heavy fragment with a
moderately deformed light fragment. SL consists of two elongated fragments in a symmetric
configuration. The AS1 mass split, $(A_L,A_H)=(99,137)$, and the AS2 mass split,
$(A_L,A_H)=(96,140)$, are consistent with those associated with the conventional
Standard~I and Standard~II modes, respectively. However, the average TKE is higher for
AS2 (177.3~MeV) than for AS1 (164.4~MeV), contrary to the conventional picture of
Standard~I and Standard~II. This TKE ordering reflects the classification of the present
asymmetric modes by the individual fragment deformations rather than by TKE, which is
primarily governed by the total deformation at scission.

Fragment mass--TKE distributions are calculated for each mode. AS2 extends over a broad
asymmetric mass range predominantly at high TKE, whereas AS1 extends toward lower TKE.
SL is concentrated around symmetric mass division at far lower TKE. In the
one-dimensional fragment mass and TKE distributions, the three modes overlap and are not clearly
separated, although the average TKE as a function of fragment mass differs among the
modes.

The single-particle level patterns reveal characteristic shell structures in the
representative fragments of the two asymmetric modes. The AS1 heavy fragment exhibits
proton gaps at $Z=50,52$ and neutron gaps at $N=82,84$, indicating an intermediate shell
structure with both spherical and deformed features. In the AS2 heavy fragment, the
spherical gaps at $Z=50$ and $N=82$ disappear, whereas well-developed gaps appear at
$Z=56$ and $N=88$, a pattern consistent with shell features associated with quadrupole
and octupole deformations in the $^{144}$Ba region~\cite{scamps2018}. The AS2 light
fragment exhibits shell features similar to those reported in a microscopic calculation
for $^{236}$U fission~\cite{bernard2023}. By contrast, the proton gap pattern of the AS1
light fragment is not directly accounted for by that microscopic calculation but
suggests a possible connection to the deformed proton shell at $Z=36$ identified in
studies of lighter fissioning systems, particularly Hg
isotopes~\cite{scamps2019,morfouace2025}.

The representative SL fragment exhibits a pronounced proton gap at $Z=46$. The SL mode
has traditionally been interpreted in terms of macroscopic liquid-drop effects. The
proton gap suggests a contribution from proton shell effects to the stability of the
elongated SL configuration. The distinct low-energy region near symmetric mass division
in the PES (Fig.~\ref{fig:pot}) is consistent with this interpretation.

The classification based on fragment mass and the quadrupole deformations of both
fragments opens a new microscopic interpretation of fission modes by identifying the
fragment shell structures associated with each mode. Experimental mass--TKE
distributions for a wider range of fissioning nuclei are needed to assess the
correspondence between this classification and phenomenological fission modes, because
the different modes overlap extensively in the fragment mass and TKE distributions.
Recent measurements of $^{258}$Md fission~\cite{nishio2025} and a comparative Langevin
study using the same classification~\cite{okada2026md} provide a concrete case in which
this correspondence can be examined. Applying the classification to other systems and
examining the resulting mode properties could help clarify the fission paths associated
with each mode.

\begin{acknowledgments}
This work was supported by JSPS KAKENHI Grant Number JP24K22887.
The JAEA supercomputer HPE SGI8600 was used for the six-dimensional Langevin
calculations, mode classification, and all other computationally intensive
tasks.
\end{acknowledgments}

\appendix
\section{Details of the $k$-means mode classification}
\label{app:kmeans}

This appendix gives the details of the $k$-means mode classification
introduced in Sec.~\ref{sec:kmeans}. At scission, each Langevin event is
specified by the six Cassini parameters. For the classification, these
parameters are used to construct two data points for each event, each
representing one of the fragments, in the three-dimensional
$(A_F,\beta_{2,F},\beta_{2,\bar F})$ space. Here, $F$ denotes the fragment
represented by a given data point and $\bar F$ its partner. Because TKE is
strongly related to the fragment deformations, it is not treated as an
independent classification variable and is examined only after the mode
assignment. The octupole deformation $\beta_3$ is also excluded, despite its
importance as a shape degree of freedom. Including additional deformation
variables would reduce the relative contribution of fragment mass, the
primary variable for fission mode classification.

Specifically, the two data points for each scission event $e$ are defined as
\begin{align}
  \bm{y}_e^{(L)}
  &=
  (A_L,\beta_{2,L},\beta_{2,H}),
  \nonumber\\
  \bm{y}_e^{(H)}
  &=
  (A_H,\beta_{2,H},\beta_{2,L}).
\end{align}
The data points $\bm{y}_e^{(L)}$ and $\bm{y}_e^{(H)}$ represent the light and heavy fragment sides, respectively.
This registration treats the two fragment sides symmetrically.
In the following, the registered points are indexed by $n$.

The three components are standardized before clustering because the
$k$-means algorithm uses Euclidean distances. The standardized coordinates
are defined by
\begin{align}
  x_{n,i}
  &=
  \frac{y_{n,i}-\bar y_i}{s_i},
  \qquad
  i=1,2,3,
\end{align}
where $\bar y_i$ and $s_i$ are the mean value and standard deviation of the
$i$th component over all registered points. The resulting standardized data
point is denoted by $\bm{x}_n$.

The number of clusters in the $k$-means classification is set to $K=5$.
Each fission event is represented by one data point for the light fragment
and another for the heavy fragment. Each asymmetric physical mode therefore
appears as a pair of mirror clusters in the registered data set. Two numerical
clusters are required for each of AS1 and AS2, whereas the symmetric elongated
component is represented by one central cluster near symmetric mass division.
After convergence, the five numerical clusters are identified from their
centroid positions and combined into the three physical modes labeled AS1,
AS2, and SL.

The clusters are obtained by minimizing the within-cluster sum of squared
distances,
\begin{align}
  W
  &=
  \sum_{c=1}^{K}
  \sum_{n\in C_c}
  \left\|
  \bm{x}_n-\bm{\mu}_c
  \right\|^2 ,
\end{align}
where $C_c$ is the set of data points assigned to cluster $c$ and
$\bm{\mu}_c$ is the corresponding centroid in standardized space. For fixed
centroids, each data point is assigned to the nearest centroid,
\begin{align}
  c(n)
  &=
  \arg\min_c
  \left\|
  \bm{x}_n-\bm{\mu}_c
  \right\|^2 ,
\end{align}
and the centroid of each cluster is updated as
\begin{align}
  \bm{\mu}_c
  &=
  \frac{1}{N_c}
  \sum_{n\in C_c}
  \bm{x}_n .
\end{align}
Here, $N_c$ is the number of data points in $C_c$. The assignment and
centroid update steps are repeated until the centroids no longer change.

\begin{figure}[t]
  \centering
  \includegraphics[width=\columnwidth, trim=0.70cm 0.75cm 0.68cm 0.58cm, clip]{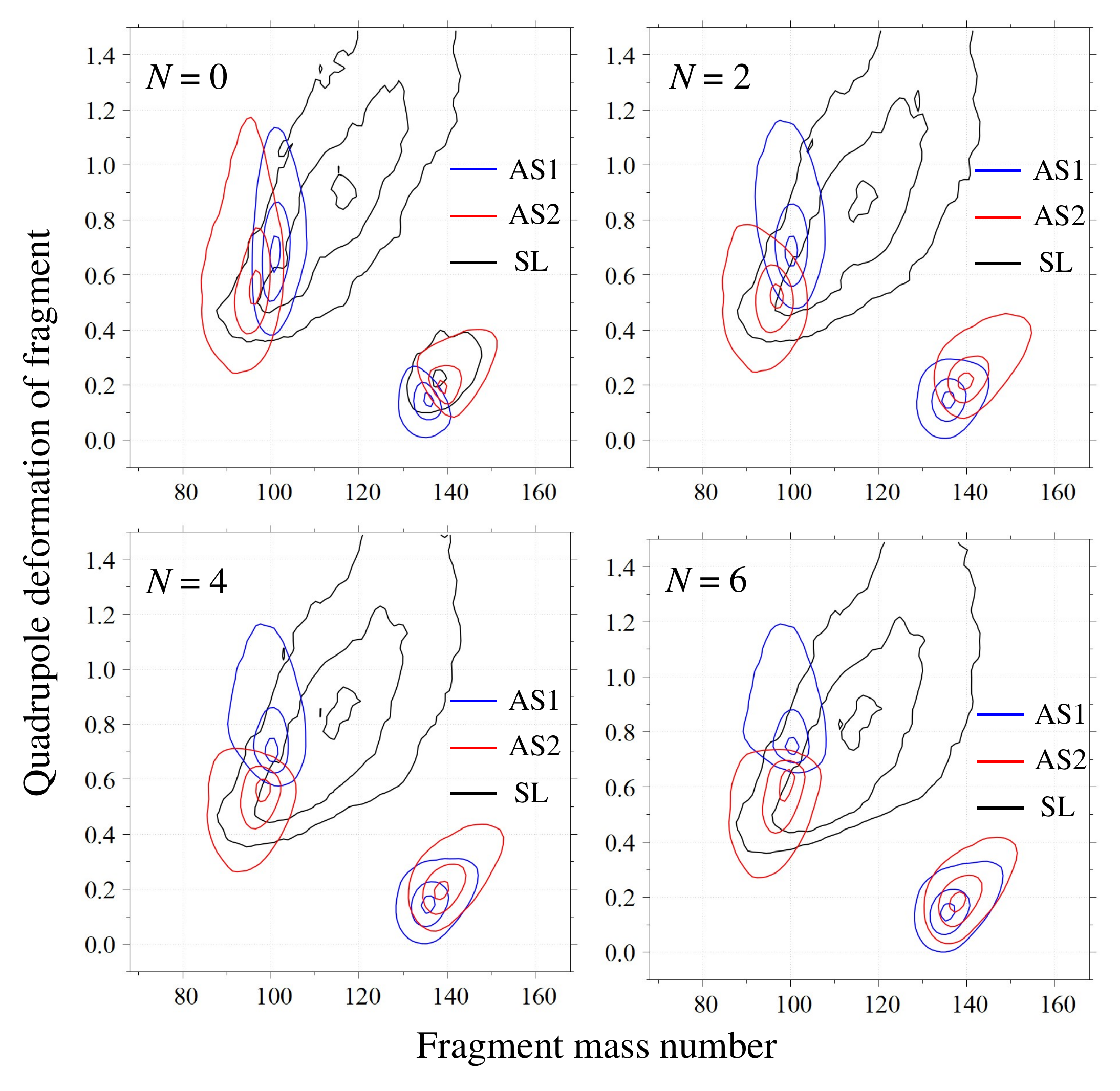}
  \caption{
  Evolution of the mode contours in the $(A_F,\beta_{2,F})$ plane. The panels
  show the contours at iterations $N=0$, 2, 4, and 6. The blue and red contours
  denote the mirror cluster pairs assigned to AS1 and AS2, respectively, and the
  black contours denote the central cluster assigned to SL. All contours are
  two-dimensional projections of the three-dimensional classification result
  and are drawn at 10\%, 50\%, and 90\% of the maximum projected density for
  each mode.
  }
  \label{fig:b2fmdi}
\end{figure}

Figure~\ref{fig:b2fmdi} shows how the projected mode contours change during
the iterations. Here, $N$ denotes the iteration number. In the initial
assignment shown in the panel labeled $N=0$, two centroid pairs are placed
near the mass distribution peaks corresponding to the light and heavy
fragments. The members of each pair differ only slightly in $A_F$, and the
remaining centroid is placed near symmetric mass division. The contours at
$N=0$ therefore mainly reflect the centroid placement based on fragment mass.
As the assignment and centroid update steps are repeated, the centroids move
in the standardized three-dimensional space
$(A_F,\beta_{2,F},\beta_{2,\bar F})$, and the projected contours gradually
come to reflect the deformation-dependent structure of the distribution.
The small changes in the projected AS1, AS2, and SL contours between $N=4$
and $N=6$ suggest that the iterative procedure is approaching convergence.
For the final mode assignment, the iterations are continued until the
centroids no longer change. For visualization, the contours in
Fig.~\ref{fig:b2fmdi} are shown as two-dimensional projections onto the
$(A_F,\beta_{2,F})$ plane, whereas the classification itself is performed in
the three-dimensional $(A_F,\beta_{2,F},\beta_{2,\bar F})$ space.

The dependence on the initial centroid positions was examined by repeating
the calculation with several centroid sets placed around the main density
regions. After the two mirror cluster pairs and the central cluster were
assigned to AS1, AS2, and SL, respectively, the resulting mode classification
was unchanged for all initial centroid sets considered.

\section{Results for individual fission modes in 14~MeV neutron-induced fission of $^{235}$U}
\label{app:n14}

This appendix examines whether the main mode characteristics persist at higher
excitation energy in 14~MeV neutron-induced fission of $^{235}$U. The mode
classification procedure described in Sec.~\ref{sec:kmeans} is applied to the
14~MeV calculation independently of the classification for thermal
neutron-induced fission.

\begin{table}[h]
  \centering
  \caption{Representative scission configurations and associated physical quantities
  for 14~MeV neutron-induced fission of $^{235}$U. For each mode, the fragment
  mass numbers are indicated inside the representative shape, and the
  quadrupole and octupole deformations of each fragment, TKE, and yield are
  listed.}
  \label{tab:n14_mode_average}
  \setlength{\tabcolsep}{4pt}
  \begin{tabular}{ l|ccc }
    \hline
    Mode & AS1 & AS2 & SL \\
    \hline
    \raisebox{2.5ex}{Shape} \rule{0pt}{6.5ex}
    & \modeshapegraphic{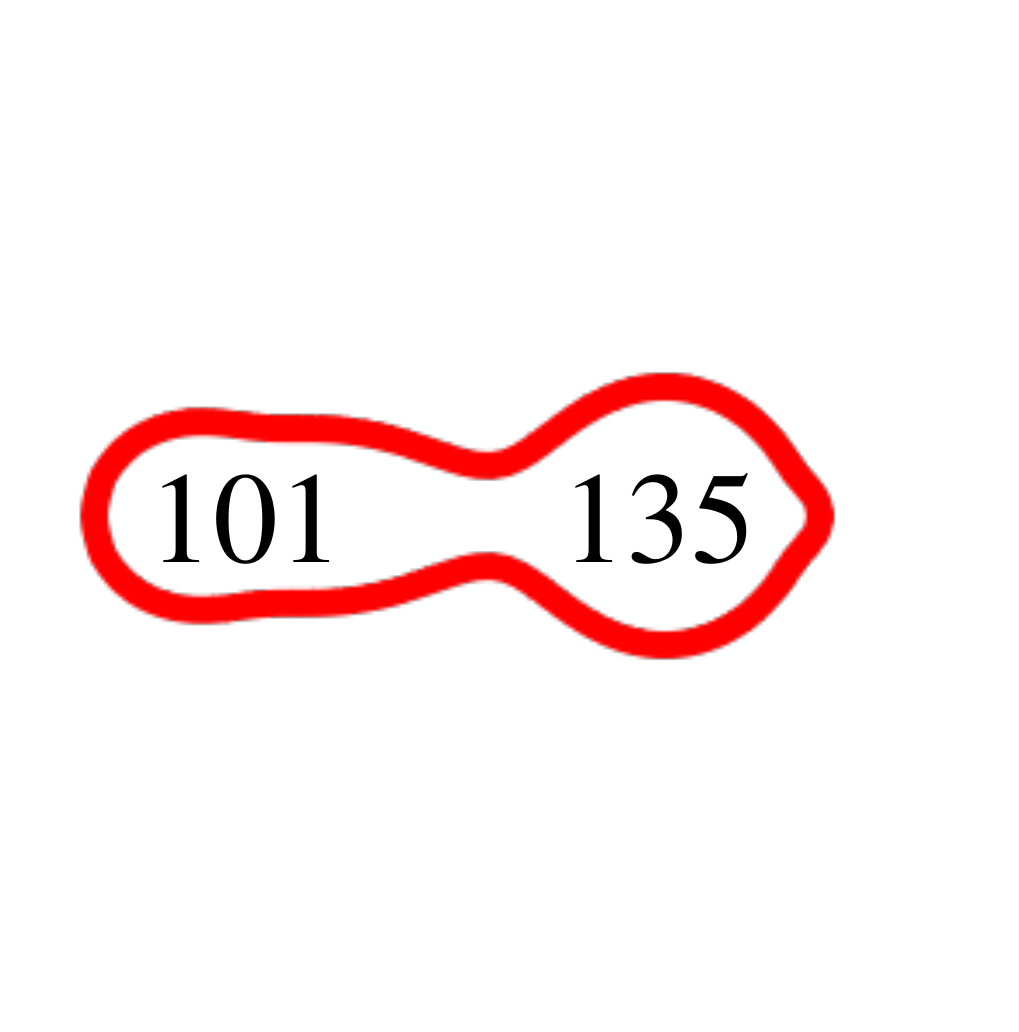}
    & \modeshapegraphic{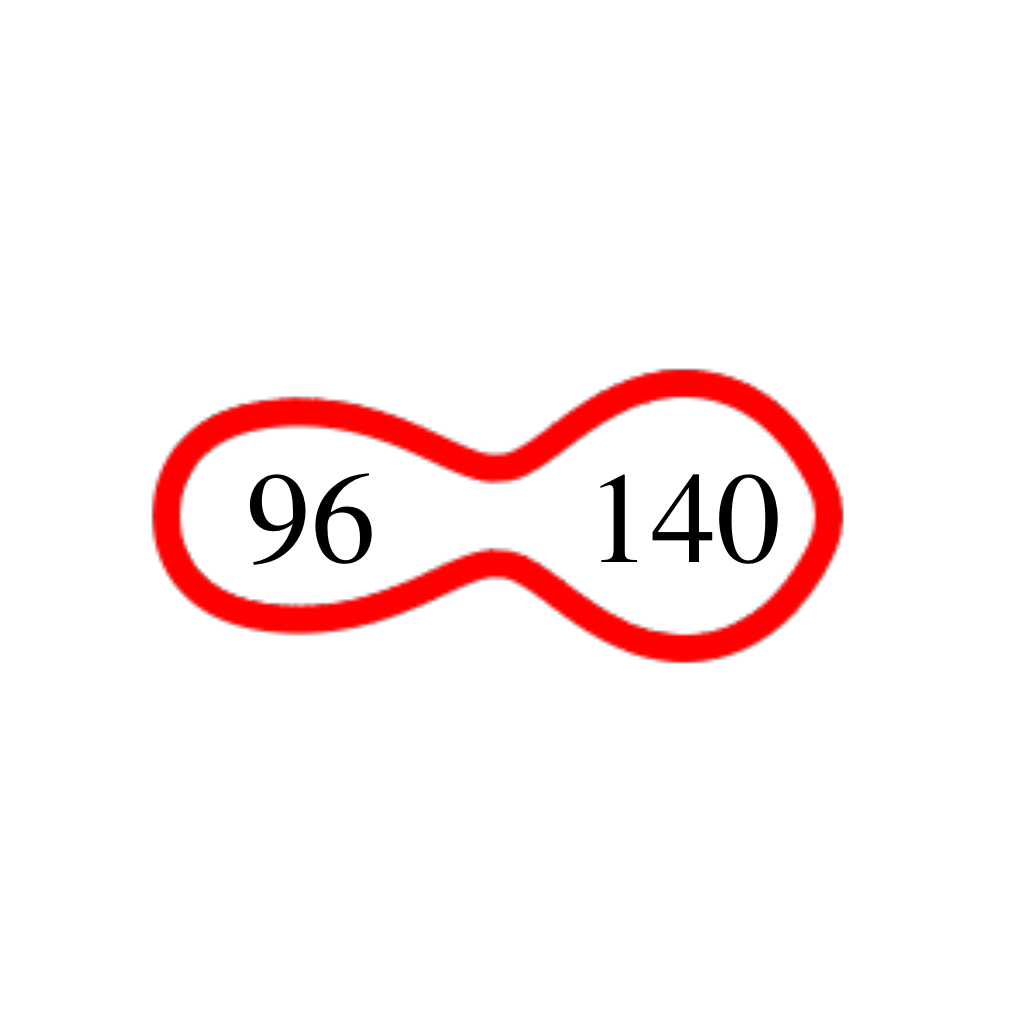}
    & \modeshapegraphic{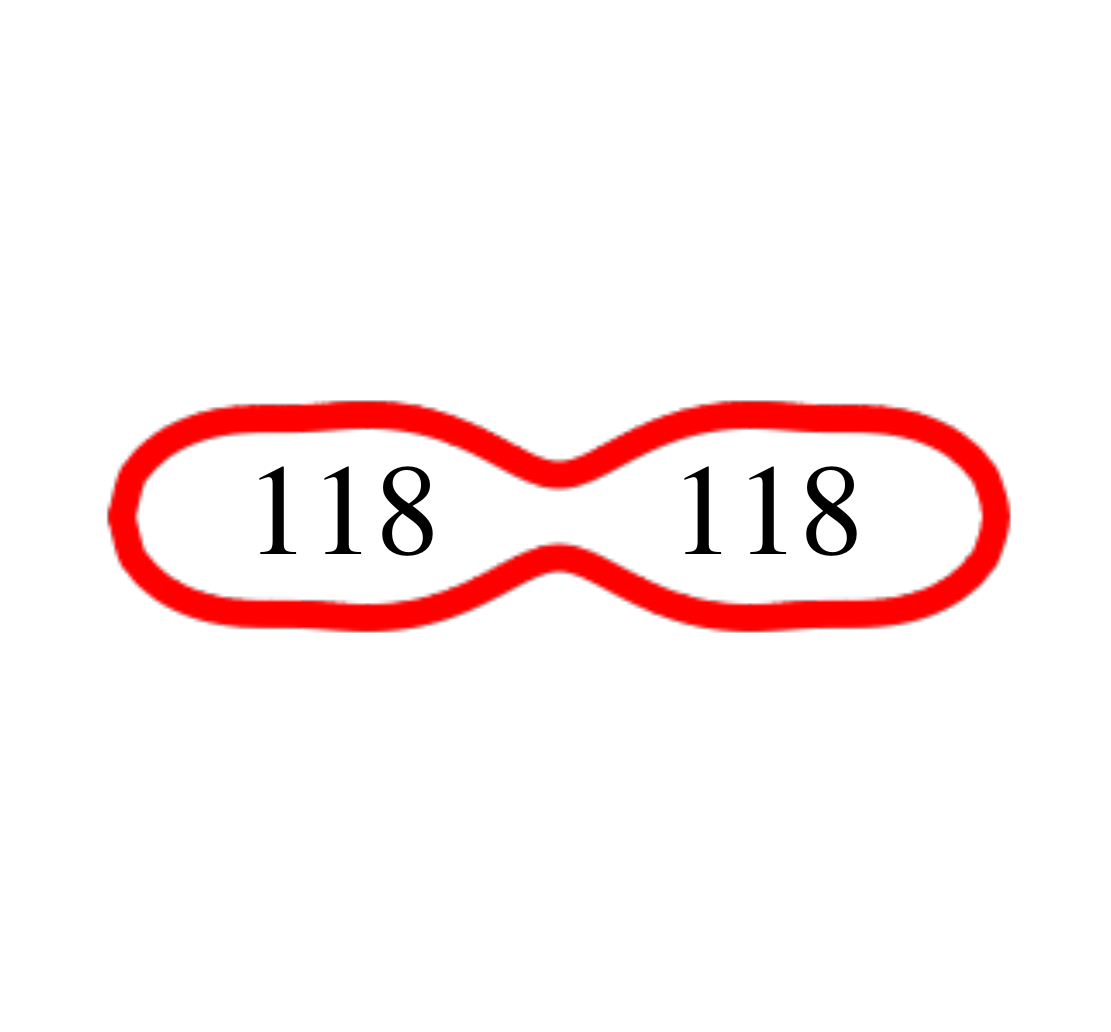} \\
    \hline
    $\beta_{2,L}$, $\beta_{2,H}$ & 1.26, 0.27 & 0.62, 0.29 & 1.01, 1.01 \\
    $\beta_{3,L}$, $\beta_{3,H}$ & 0.46, 0.13 & 0.31, 0.18 & 0.15, 0.15 \\
    TKE (MeV)           & 159.7      & 179.0      & 153.1 \\
    Yield (\%)          & 23.3       & 69.5       & 7.2 \\
    \hline
  \end{tabular}
\end{table}

Table~\ref{tab:n14_mode_average} summarizes the mode-averaged quantities for
the 14~MeV calculation. The AS1 mass split becomes slightly less asymmetric,
changing from $A_L/A_H=99/137$ for thermal neutron-induced fission to
$101/135$ for 14~MeV neutron-induced fission, whereas the AS2 and SL mass
splits remain nearly unchanged. The AS2 yield changes only slightly, from
68.6\% to 69.5\%, whereas the AS1 yield decreases from 27.1\% to 23.3\% and
the SL yield increases from 4.3\% to 7.2\%.

The quadrupole deformations of both fragments increase in all three modes. For
AS1, the deformation pair $(\beta_{2,L},\beta_{2,H})$ changes from
$(1.03,0.20)$ to $(1.26,0.27)$, whereas the corresponding AS2 pair changes
from $(0.58,0.23)$ to $(0.62,0.29)$. For SL, both fragment deformations
increase from 0.88 to 1.01. Despite these increases, the contrast between the
strongly elongated AS1 light fragment and the moderately deformed AS2 light
fragment is preserved.

\begin{figure}[t]
  \centering
  \includegraphics[width=\columnwidth, trim=1.30cm 1.35cm 0.50cm 5.30cm, clip]{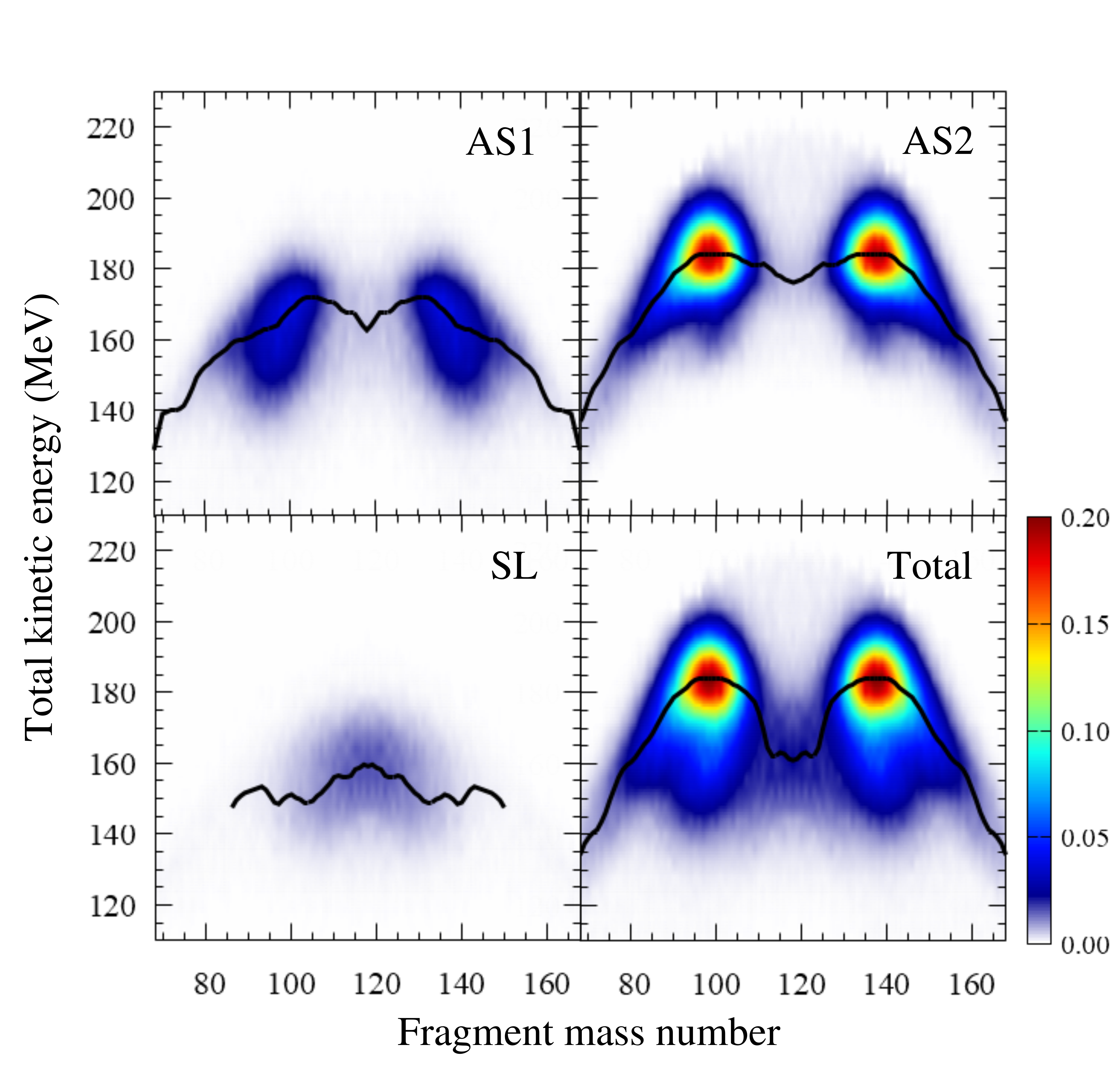}
  \caption{Mass--TKE distributions for 14~MeV neutron-induced fission of $^{235}$U.
  The upper-left, upper-right, lower-left, and lower-right panels show the AS1, AS2,
  SL, and total distributions, respectively. The color scale represents the density
  in bins with widths $\Delta A_F=1$ and $\Delta \mathrm{TKE}=4~\mathrm{MeV}$.
  The total distribution is normalized to 200\%. The black curve in each panel
  indicates the TKE corresponding to the peak of the distribution at each fragment mass number.}
  \label{fig:TKEFMD_n14}
\end{figure}

Figure~\ref{fig:TKEFMD_n14} shows the mass--TKE distributions for each mode
in the 14~MeV calculation. The distributions are broader than those in
Fig.~\ref{fig:TKEFMD_nth}, but the general mode characteristics are
preserved. AS2 forms the high-TKE component in the asymmetric mass region,
AS1 extends to lower TKE, and SL is concentrated near symmetric mass
division.

\FloatBarrier

\begin{figure}[h]
  \centering
  \includegraphics[width=\columnwidth, trim=1.45cm 1.55cm 0.20cm 1.75cm, clip]{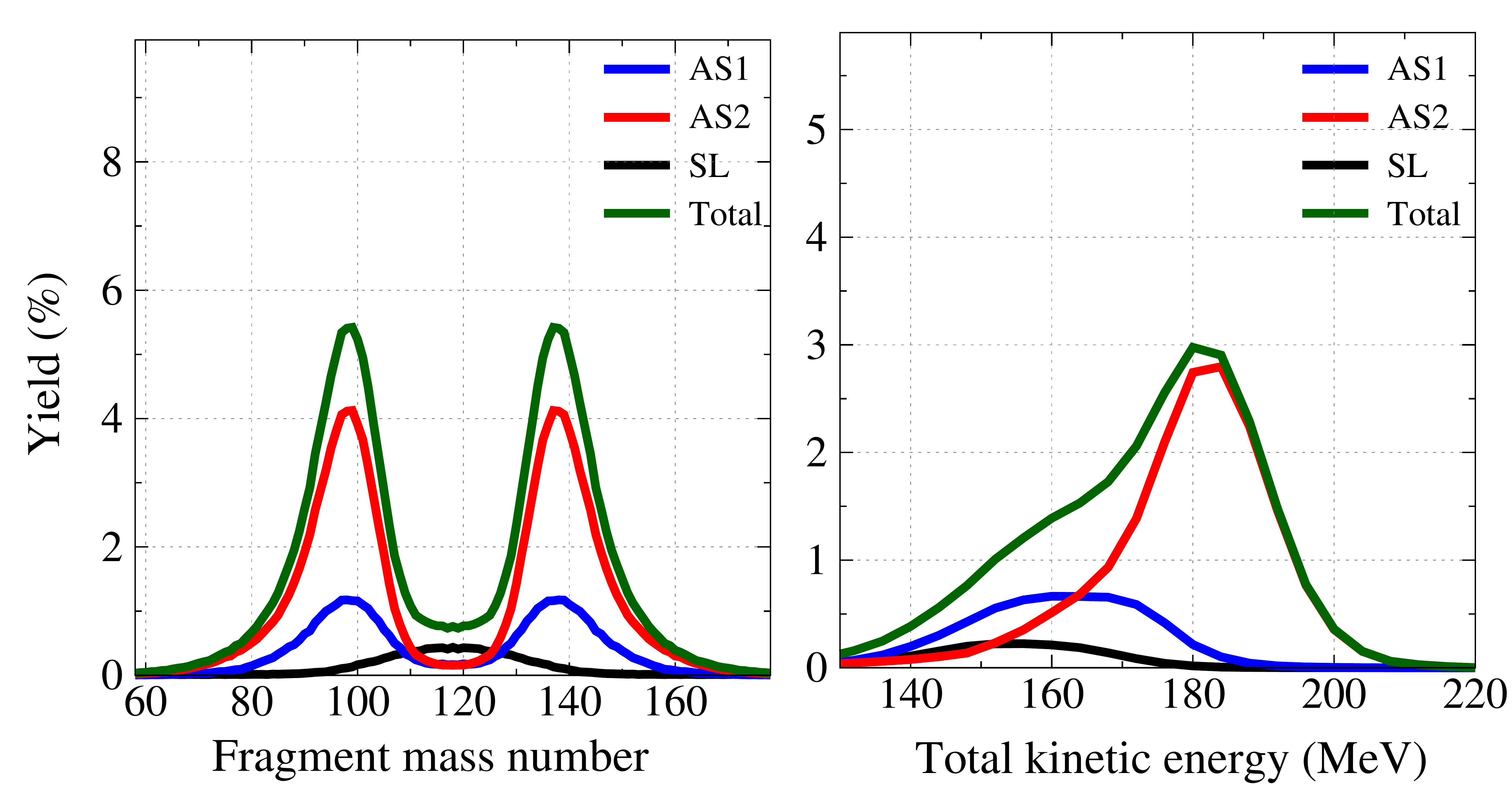}
  \caption{One-dimensional projections of the mass--TKE distributions for each mode
  in 14~MeV neutron-induced fission of $^{235}$U. The left and right panels show
  the fragment mass and TKE distributions, respectively.}
  \label{fig:mode_14}
\end{figure}

Figure~\ref{fig:mode_14} shows the corresponding one-dimensional fragment
mass and TKE distributions. The AS2 component dominates the yield in both the
asymmetric mass region and the high-TKE region, whereas the smaller AS1
contribution extends to lower TKE. The SL contribution remains smaller than
the AS1 and AS2 contributions but is more prominent than in the thermal
neutron-induced fission calculation.

A full comparison with experimental data for 14~MeV neutron-induced fission
of $^{235}$U would require extending the present $^{236}$U calculation to
include additional processes, such as prefission neutron emission and
multichance fission. Within the present calculation, however, the persistence
of the mode characteristics despite the broader distributions suggests that
the present approach can track the individual modes at the higher excitation
energy.

\bibliographystyle{apsrev4-2}
\bibliography{myref_Md}
\end{document}